# An Anytime Algorithm for Optimal Coalition Structure Generation

**Talal Rahwan** TR@ECS.SOTON.AC.UK
**Sarvapali D. Ramchurn** SDR@ECS.SOTON.AC.UK
**Nicholas R. Jennings** NRJ@ECS.SOTON.AC.UK
*School of Electronics and Computer Science,*
*University of Southampton, Southampton, SO17 1BJ, U.K.*

**Andrea Giovannucci** AGIOVANNUCCI@IUA.UPF.EDU
*SPECS Laboratory, Pompeu Fabra University, Barcelona, Spain.*

## Abstract

Coalition formation is a fundamental type of interaction that involves the creation of coherent groupings of distinct, autonomous, agents in order to efficiently achieve their individual or collective goals. Forming effective coalitions is a major research challenge in the field of multi-agent systems. Central to this endeavour is the problem of determining which of the many possible coalitions to form in order to achieve some goal. This usually requires calculating a value for every possible coalition, known as the *coalition value*, which indicates how beneficial that coalition would be if it was formed. Once these values are calculated, the agents usually need to find a combination of coalitions, in which every agent belongs to exactly one coalition, and by which the overall outcome of the system is maximized. However, this *coalition structure generation* problem is extremely challenging due to the number of possible solutions that need to be examined, which grows exponentially with the number of agents involved. To date, therefore, many algorithms have been proposed to solve this problem using different techniques — ranging from dynamic programming, to integer programming, to stochastic search — all of which suffer from major limitations relating to execution time, solution quality, and memory requirements.

With this in mind, we develop an anytime algorithm to solve the coalition structure generation problem. Specifically, the algorithm uses a novel representation of the search space, which partitions the space of possible solutions into sub-spaces such that it is possible to compute upper and lower bounds on the values of the best coalition structures in them. These bounds are then used to identify the sub-spaces that have no potential of containing the optimal solution so that they can be pruned. The algorithm, then, searches through the remaining sub-spaces very efficiently using a branch-and-bound technique to avoid examining all the solutions within the searched sub-space(s). In this setting, we prove that our algorithm enumerates all coalition structures efficiently by avoiding redundant and invalid solutions automatically. Moreover, in order to effectively test our algorithm we develop a new type of input distribution which allows us to generate more reliable benchmarks compared to the input distributions previously used in the field. Given this new distribution, we show that for 27 agents our algorithm is able to find solutions that are optimal in $0.175\%$ of the time required by the fastest available algorithm in the literature. The algorithm is anytime, and if interrupted before it would have normally terminated, it can still provide a solution that is guaranteed to be within a bound from the optimal one. Moreover, the guarantees we provide on the quality of the solution are significantly better than those provided by the previous state of the art algorithms designed for this purpose. For example, for the worst case distribution given 25 agents, our algorithm is able to find a 90% efficient solution in around 10% of time it takes to find the optimal solution.

## 1. Introduction

Multi-agent systems are considered an important and rapidly expanding area of research in artificial intelligence. This is due to its natural fit to many real-world scenarios, and its wide variety of applications (Jennings, 2001). Now, typically, the agents in a multi-agent system need to be organized such that the roles, relationships, and authority structures that govern their behaviour are clearly defined (Horling & Lesser, 2005). Different organizational paradigms include hierarchies, teams, federations, and many others, each with its own strengths and weaknesses, making it more suitable for some problems, and less suitable for others. Among the organizational paradigms that are becoming increasingly important is that of coalitions. Coalitions can be distinguished from other organizations by being *goal-directed and short-lived*; i.e. the coalitions are formed with a purpose in mind, and are dissolved when that purpose no longer exists, or when they cease to suit their designed purpose, or when the profitability is lost as agents depart (Horling & Lesser, 2005). Another defining feature is that within each coalition, the agents coordinate their activities in order to achieve the coalition's goal(s), but no coordination takes place among agents belonging to different coalitions (except if the coalitions' goals interact). Moreover, the organizational structure within each coalition is usually flat (although there could be a coalition leader acting as a representative for the group as a whole).

The area of coalition formation has received considerable attention in recent research, and has proved to be useful in a number of real-world scenarios and multi-agent systems. For example, in e-commerce, buyers can form coalitions to purchase a product in bulk and take advantage of price discounts (Tsvetovat, Sycara, Chen, & Ying, 2000). In e-business, groups of agents can be formed in order to satisfy particular market niches (Norman, Preece, Chalmers, Jennings, Luck, Dang, Nguyen, V. Deora, Gray, & Fiddian, 2004). In distributed sensor networks, coalitions of sensors can work together to track targets of interest (Dang, Dash, Rogers, & Jennings, 2006). In distributed vehicle routing, coalitions of delivery companies can be formed to reduce the transportation costs by sharing deliveries (Sandholm & Lesser, 1997). Coalition formation can also be used for information gathering, where several information servers form coalitions to answer queries (Klusch & Shehory, 1996).

Generally speaking, the coalition formation process can be viewed as being composed of the three main activities that are outlined below (Sandholm, Larson, Andersson, Shehory, & Tohme, 1999):

1. **Coalition Value Calculation:** In this context, a number of coalition formation algorithms have been developed to determine which of the potential coalitions should actually be formed. To do so, they typically calculate a value for each coalition, known as the *coalition value*, which provides an indication of the expected outcome that could be derived if that coalition was formed. Then, having computed all the coalition values, the decision about the optimal coalition(s) to form can be taken. The way this value is calculated depends on the problem under investigation.

   In an electronic marketplace, for example, the value of a coalition of buyers can be calculated as the difference between the sum of the reservation costs of the coalition members and the minimum cost needed to satisfy the requests of all the members (Li & Sycara, 2002). In information gathering, the coalition value can be designed to represent a measure of how closely the information agents' domains are related (Klusch & Shehory, 1996). In cases where the agents' rationality is bounded due to computational complexity, the value of a coalition may represent the best outcome it can achieve given limited computational resources for solving the problem (Sandholm & Lesser, 1997).

2. **Coalition Structure Generation:** Having computed the coalition values, the coalition structure generation (CSG) problem involves partitioning the set of agents into exhaustive and disjoint coalitions so as to maximize social welfare. Such a partition is called a *coalition structure*. For example, given a set of agents $A = \{a_1, a_2, a_3\}$, there exist five possible coalition structures: $\{\{a_1\}, \{a_2\}, \{a_3\}\}$, $\{\{a_1\}, \{a_2, a_3\}\}$, $\{\{a_2\}, \{a_1, a_3\}\}$, $\{\{a_3\}, \{a_1, a_2\}\}$, $\{\{a_1, a_2, a_3\}\}$.

   It is usually assumed that every coalition performs equally well, given any coalition structure containing it (i.e. the value of a coalition does not depend on the actions of non-members). Such settings are known as *characteristic function games (CFGs)*, where the value of a coalition is given by a *characteristic function*. Many, but clearly not all, real-world multi-agent problems happen to be CFGs (Sandholm et al., 1999).

   Note that an optimal solution to the CSG problem is one that maximizes the social welfare. Now, unlike a cooperative environment where the agents are mainly concerned with maximizing the social welfare, the agents in a selfish environment are only concerned with maximizing their own utility. This, however, does not mean that a CSG algorithm cannot be applied in selfish multi-agent systems. This is because the designer of such systems is usually concerned with raising the overall efficiency of the system and, in many cases, this corresponds to maximizing the social welfare. To this end, the designer needs to design an enforcement mechanism that motivates the agents to join the optimal coalition structure and, in order to do so, he first needs to know what that structure is. Moreover, knowing the value of the optimal coalition structure, or knowing a value that is within a bound from that optimal, allows the designer to evaluate the relative effectiveness of the coalition structure currently formed in the system.

3. **Pay-off Distribution:** Having determined which coalitions should be formed, it is important to determine the rewards that each agent should get in order for the coalitions to be *stable*. Here, stability refers to the state where the agents have no incentive to deviate from the coalitions to which they belong (or little incentive in weaker types of stability). This is desirable because it ensures that the agents will devote their resources to their chosen coalition rather than negotiating with, and moving to, other coalitions. This ensures that the coalitions can last long enough to actually achieve their goals. The analysis of such incentives has long been studied within the realm of *cooperative game theory*. In this context, many solutions have been proposed based on different stability concepts. These include the *Core*, the *Shapley value*, and the *Kernel* (more details can be found in the paper by Osborne & Rubinstein, 1994). Moreover, schemes have been developed to transfer non-stable pay-off distributions to stable ones while keeping the coalition structure unchanged (Kahan & Rapoport, 1984, provide a comprehensive review on stability concepts and transfer schemes in game theory). Note, however, that the agents in a cooperative environment have no incentive to dissolve a coalition that improves the performance of the system as a whole. Therefore, pay-off distribution is less important, and the main concern is generating a coalition structure so as to maximize the social welfare.

One of the most challenging of all of these activities is that of coalition structure generation, and this is due to the number of possible solutions which grows exponentially (in $O(n^n)$ and $\omega(n^{\frac{n}{2}})$ with the number of agents involved ($n$)). More specifically, it has been proved that finding an optimal coalition structure is NP-complete (Sandholm et al., 1999). To combat this complexity, a number of algorithms have been developed in the past few years, using different search techniques (e.g. dynamic programming, integer programming, and stochastic search). These algorithms, however, suffer from major limitations that make them either inefficient or inapplicable, particularly given larger numbers of agents (see Section 2 for more details).

This motivates our aim to develop an efficient algorithm for searching the space of possible coalition structures. In more detail, given a CFG setting, we wish to develop an algorithm that satisfies the following properties:

1. *Optimality*: When run to completion, the algorithm must always be able return a solution that maximizes the social welfare.

2. *Ability to prune*: the algorithm must be able to identify the sub-spaces that have no potential of containing an optimal solution so that they can be pruned from the search space. This property is critical given the exponential nature of the problem (e.g. given 20 agents, the number of possible coalition structures is 51,724,158,235,372).

3. *Discrimination*: the algorithm must be able to verify, during the search, that it has found an optimal solution, instead of proceeding with the search in the hope that a better solution can be found.

4. *Anytime*: the algorithm should be able to quickly return an initial solution, and then improve on the quality of this solution as it searches more of the space, until it finds an optimal one. This is particularly important since the agents might not always have sufficient time to run the algorithm to completion, especially given the exponential size of the search space. Moreover, being anytime makes the algorithm more robust against failure; if the execution is stopped before the algorithm would have normally terminated, then it would still provide the agents with a solution that is better than the initial solution, or any other intermediate one.

5. *Worst Case Guarantees*: the algorithm should be able to provide worst-case guarantees on the quality of its solution. Otherwise, the generated solution could always be arbitrarily worse than the optimal one. Such guarantees are important when trading off between the solution quality and the search time. For example, if the quality of the current solution is known to be no worse than, say, $95\%$ of the optimal one, and if there is still a significant portion of the space left to be searched, then the agents might decide that it is not worthwhile to carry on with the search. Obviously, the better the guarantees, the more likely it is that the agents will decide to stop searching for a better solution.

Against the research aims outlined above, this paper makes the following contributions to the state of the art in coalition structure generation:

1. We provide a new representation of the space of possible coalition structures. This representation partitions the space into much smaller, disjoint sub-spaces that can be explored independently to find an optimal solution. As opposed to the other widely-used representation (Sandholm et al., 1999; Dang & Jennings, 2004), by which the coalition structures are categorized based on the *number of coalitions* they contain, our representation categorizes the coalition structures into sub-spaces based on the *sizes of the coalitions* they contain. A key advantage of this representation is that, immediately after scanning the input to the algorithm (i.e. the coalition values), we can compute the average value of the coalition structures within each sub-space. Moreover, by scanning the input, we can also compute an upper and a lower bound on the value of the best coalition structure that could be found in each of these sub-spaces. Then, by comparing these bounds, it is possible to identify the sub-spaces that have no potential of containing an optimal solution so that they can be pruned. A second major advantage of this representation is that it allows the agents to analyse the trade-off between the size of (i.e. the number of coalition structures within) a sub-space and the improvement it may bring to the current solution by virtue of its bounds. Hence, rather than constraining the solution to fixed sizes, as Shehory and Kraus (1998) do, agents using our representation can make a more informed decision about the sizes of coalitions to choose (since each of the sub-spaces are defined by the sizes of coalitions within the coalition structures).

2. We develop a novel, anytime, integer-partition based algorithm (called IP) for coalitions structure generation which uses the representation discussed above, and provides very high guarantees on the quality of its solutions very quickly. Moreover, IP is guaranteed to return an optimal solution when run to completion.

3. We prove that our algorithm is able to enumerate coalition structures efficiently by avoiding redundant and invalid solutions. Our enumeration technique also allows us to apply branch-and-bound to reduce the amount of search needed.

4. While many CSG algorithms in the literature have been evaluated using the input distributions that were defined by Larson and Sandholm (2000), we prove that these distributions are biased as far as the CSG problem is concerned. Moreover, we propose a new distribution and prove that it tackles this problem, making it much more suitable for evaluating CSG algorithms in general.

5. When evaluating the time required to return an optimal solution, we compare IP with the fastest algorithm guaranteed to return an optimal solution (i.e. the Improved Dynamic Programming (IDP) algorithm by Rahwan & Jennings, 2008b). This comparison shows that IP is significantly faster. In more detail, IP is empirically shown to find an optimal solution in $0.175\%$ of the time taken by IDP given 27 agents.

6. We benchmark IP against previous anytime algorithms (Sandholm et al., 1999; Dang & Jennings, 2004), and show that it provides significantly better guarantees on the quality of the solutions it generates over time. In more detail, we empirically show that, for various numbers of agents, the quality of its initial solution (i.e. the solution found after scanning the input) is usually guaranteed to be at least $40\%$ of the optimal, as opposed to $\frac{2}{n}$ (which means for example, $10\%$ for 20 agents and $8\%$ for 25 agents) for both Sandholm et al.'s algorithm and Dang and Jennings's algorithm. For the standard distributions with which we evaluate our algorithm, we also find that it usually terminates by searching only minute portions of the search space and generates near-optimal solutions (i.e. $> 90\%$ of the optimal) by searching even smaller portions of the search space (i.e. on average around $0.0000002\%$ of the search space). This is a tremendous improvement over the aforementioned algorithms which could guarantee solutions higher than $50\%$ of the optimal only after searching the whole space.

Note that this is a significantly revised and extended version of previous papers (Rahwan, Ramchurn, Dang, & Jennings, 2007a; Rahwan, Ramchurn, Giovannucci, Dang, & Jennings, 2007b). Specifically, we provide in this paper a more comprehensive review of the available algorithms in the CSG literature. We also provide a detailed analysis of our IP algorithm, describe the pseudo code of all the functions used in IP, and prove the correctness of the function that searches the different sub-spaces. A mathematical proof is also provided regarding the way the size of a sub-space is computed. Moreover, we question the validity of the standard value distributions that are used in the literature, and propose a new value distribution (called NDCS) that is more suitable for evaluating CSG algorithms. Finally, we benchmark our algorithm against the improved dynamic programming algorithm (IDP) by Rahwan and Jennings (2008b) (instead of the standard DP algorithm).

The remainder of the paper is organized as follows. In Section 2, we describe the algorithms that are currently available for solving the coalition structure generation problem, and discuss their relative advantages and limitations. In Section 3, we present our novel representation of the search space and, in Section 4, we present our integer-partition based algorithm (IP), showing how it identifies the sub-spaces that can be pruned, and how it searches through the remaining ones without going through invalid or redundant coalition structures, using a branch-and-bound technique. Section 5 provides an empirical evaluation of the algorithm, and benchmarks it against the current state of the art in the CSG literature. Section 6 concludes the paper and outlines future work. We also provide, in the appendices, a summary of the main notations employed, as well as detailed proofs of the theorems provided in the paper.

## 2. Related Work

Previous algorithms that have been designed for the coalition structure generation problem can be classified into two main categories:

- Exact algorithms[1] – using heuristics, integer programming, or dynamic programming.
- Non-exact algorithms – using genetic algorithms, or limiting the search space in some way.

Next, we discuss both the advantages and the limitations of the algorithms that fall within each of these classes. Throughout the paper, we denote by $n$ the number of agents, and by $A = \{a_1, a_2, \cdots, a_n\}$ the set of agents. Moreover, we define an order over the agents in $A$ as follows: $\forall a_i, a_j \in A,\ a_i < a_j$ iff $i < j$, and $a_i = a_j$ iff $i = j$. In other words, we have: $a_1 < a_2 < \cdots < a_n$. Finally, we denote by $v(C)$ the value of coalition $C$, and $V(CS)$ the value of coalition structure $CS$.

### 2.1 Exact Algorithms for Coalition Structure Generation

There are very few exact algorithms for coalition structure generation. Those that have been developed can be distinguished based on whether they use dynamic programming or heuristics. In what

1. Recall that an exact algorithm is one that always returns an optimal solution if it exists (Evans & Minieka, 1992).

follows, we outline their features and discuss how they relate to our ultimate goal of developing an efficient, anytime, optimal coalition structure generation algorithm.

#### 2.1.1 DYNAMIC PROGRAMMING

Here we consider computationally efficient algorithms designed to return an optimal solution. Note that the emphasis, here, is on providing a guarantee on the performance of the algorithm in worst-case scenarios. In this context, Yeh (1986) developed a dynamic programming algorithm to solve the complete set partitioning problem. A very similar algorithm was later developed by Rothkopf, Pekec, and Harstad (1995) to solve the winner determination problem in combinatorial auctions. These algorithms can be directly applied to find optimal coalition structures, since the problems they were originally designed to solve are very similar to the CSG problem.[2] Also note that both of these algorithms use basically the same technique and, therefore, have the same computational complexity. Thus, throughout this paper, we do not distinguish between them, and refer to both as the dynamic programming (DP) algorithm. The biggest advantage of this algorithm is that it runs in $O(3^n)$ time (Rothkopf et al., 1995). This is significantly less than exhaustive enumeration of all coalition structures (which is $O(n^n)$). In fact, DP is polynomial in the size of the input. This is because the input includes $2^n - 1$ values, and the following holds:

$$O(3^n) = O(2^{(log_2 3)n}) = O((2^n)^{log_2 3})$$

Therefore, the computational complexity of the algorithm is $O(y^{log_2 3})$, where $y$ is the number of values in the input. While, on the one hand, no other algorithm in the literature is guaranteed to find an optimal coalition structure in polynomial time (in the size of the input), on the other hand, the main limitation of DP is that it does not generate solutions anytime, and has a large memory requirement. Specifically, it requires maintaining three tables in memory containing $2^n$ entries each.

More recently, Rahwan and Jennings (2008b) developed an Improved Dynamic Programming algorithm (called IDP) that performs fewer operations and requires less memory than DP (e.g. given 25 agents, it performs only $38.7\%$ of the operations, and requires $66.6\%$ of the memory in the worst case, and $33.3\%$ in the best). However, IDP does not return solutions anytime. As mentioned earlier, this is undesirable, especially given large numbers of agents, because the time required to return the optimal solution might be longer than the time available to the agents.

#### 2.1.2 ANYTIME ALGORITHMS WITH WORST CASE GUARANTEES

Sandholm et al. (1999) were the first to introduce an anytime algorithm for coalition structure generation that establishes bounds on the quality of the solution found so far. They view the coalition structure generation process as a search in what they call the *coalition structure graph* (see Figure 1). In this undirected graph, every node represents a possible coalition structure. The nodes are categorized into $n$ levels, noted as $LV_1, \cdots, LV_n$ where level $LV_i$ contains the coalition structures that contain $i$ coalitions. The arcs represent mergers of two coalitions when followed upwards, and splits of a coalition into two coalitions when followed downwards.

2. This is because they both involve partitioning a set of elements into subsets based on the weights that are associated to every possible subset.

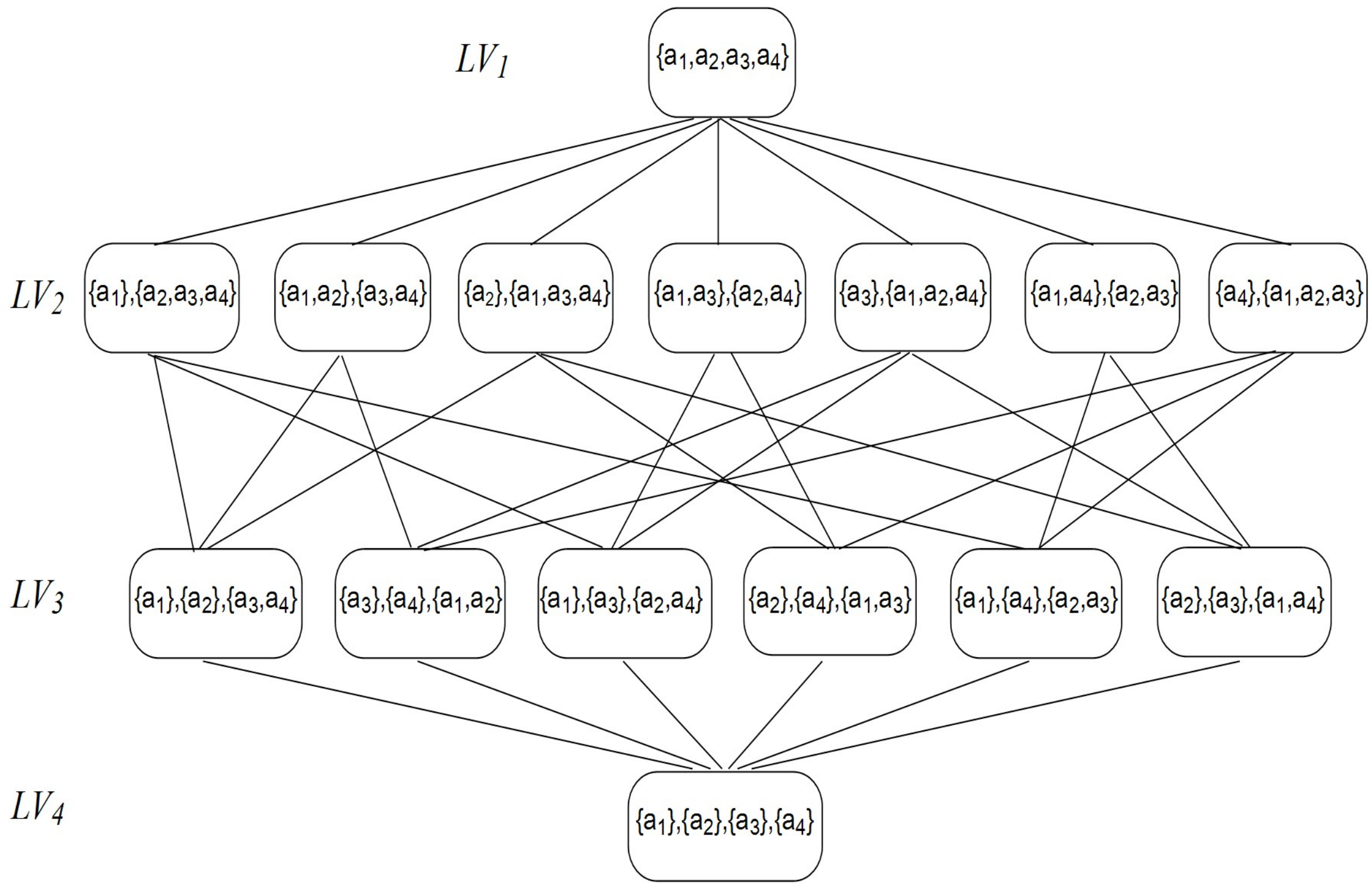

Figure 1: The coalition structure graph for 4 agents.

Sandholm et al. (1999) have proved that, in order to establish a bound on the quality of a coalition structure, it is sufficient to search through the first two levels of the coalition structure graph. In this case, the bound $\beta$ would be equal to $n$, and the number of searched coalition structures would be $2^{n-1}$. They have also proved that this bound is tight; meaning that no better bound exists for this search. Moreover, they have proved that no other search algorithm (other than the one that searches the first two levels) can establish any bound while searching only $2^{n-1}$ coalition structures or fewer. This is because, in order to establish a bound, one needs to go through a subset of coalition structures in which every coalition appears *at least once*.[3] This implies that the smallest subset of coalition structures to be searched before a bound can be established is the one in which every coalition appears *exactly once*, and the only subset in which this occurs is the one containing all the coalition structures that belong to the first two levels of the graph.

If the first two levels have been searched, and additional time remains, then it would be desirable to lower the bound with further search. Sandholm et al. (1999) have developed an algorithm for this purpose. Basically, the algorithm searches the remaining levels one by one, starting from the bottom level, and moving upwards in the graph. Moreover, Sandholm et al. have also proved that the bound $\beta$ is improved whenever the algorithm finishes searching a particular level. What is interesting here is that, by searching the bottom level (which only contains one coalition structure) the bound drops in half (i.e. $\beta = \frac{n}{2}$). Then, roughly speaking, the divisor in the bound increases by one every time

3. Otherwise, if a coalition did not appear in any of these coalition structures, and if the value of this coalition happened to be arbitrarily better than the value of other coalitions, then every coalition structure containing it would be arbitrarily better than those that do not.

two more levels are searched, but seeing only one more level helps very little (Sandholm et al., 1999).[4]

This algorithm has the advantage of being anytime, and being able to provide worst case guarantees on the quality of the solution found so far. However, the algorithm has two major limitations:

- The algorithm needs to search through *the entire search space* in order for the bound to become 1. In other words, to return a solution that is guaranteed to be optimal, the algorithm simply performs a brute-force search. As discussed in Section 1, this is intractable even for small numbers of agents.

- The bounds provided by the algorithm might be too large for practical use. For example, given $n = 24$, and given that the algorithm has finished searching levels $LV_1$, $LV_2$, and $LV_{24}$ (which contain 8,388,609 coalition structures) the bound would be $\beta = n/2 = 12$. This means that, in the worst case, the optimal solution can be 12 times better than the current solution. In other words, the value of the current solution is only guaranteed to be no worse than $8.33\%$ of the value of the optimal solution. After that, in order to reduce the bound to $\beta = n/4$, four more levels need to be searched, namely $LV_{23}$, $LV_{22}$, $LV_{21}$, and $LV_{20}$. In other words, after searching an additional 119,461,563 coalition structures, the value of the solution is only guaranteed to be no worse than $16.66\%$ of the optimal value. Similarly, to reduce the bound to $\beta = n/6$, the algorithm needs to search an additional 22,384,498,067,085 coalition structures only to guarantee that the value of the solution is no worse than $25\%$ of the optimal value. Moreover, the guarantee does not go beyond $50\%$ until the entire space has been searched.

Given the limitations of Sandholm et al.'s (1999) algorithm, Dang and Jennings (2004) developed an anytime algorithm that can also establish a bound on the quality of the solution found so far, but that uses a different search method. In more detail, the algorithm starts by searching the top two levels, as well as the bottom one (as Sandholm et al.'s algorithm does). After that, however, instead of searching through the remaining levels one by one (as Sandholm et al. do), the algorithm searches through specific subsets of the remaining levels. Figure 2 compares the performance of both algorithms, and, by looking at this figure, we can see that neither of the two algorithms significantly outperforms the other.

Note, however, that both algorithms were not meant for the case where the entire space will eventually be searched. This is because if we had enough time to perform this search, then we would have used the dynamic programming algorithm, which performs this search much quicker. Instead, these algorithms were mainly developed for the cases where the space is too large to be fully searched, even when the dynamic programming algorithm is being used.

Having discussed two algorithms that use similar techniques (i.e. by Sandholm et al., 1999 and Dang & Jennings, 2004), we now discuss a different approach that can also provide solutions anytime, and can establish worst-case guarantees on the quality of its solution. This involves the use of standard problem solving techniques that rely on general purpose solvers. In more detail, the coalition structure generation problem can be formulated as a *binary* integer programming problem (or

4. To be more precise, depending on the number of agents and the level searched, the bound will either be $\lceil \frac{n}{m} \rceil$ or $\lfloor \frac{n}{m} \rfloor$ where $m = 2, 3, \cdots, n$. However, to ease the discussion and without loss of generality, we will assume throughout the paper that the bound is simply $\frac{n}{m}$.

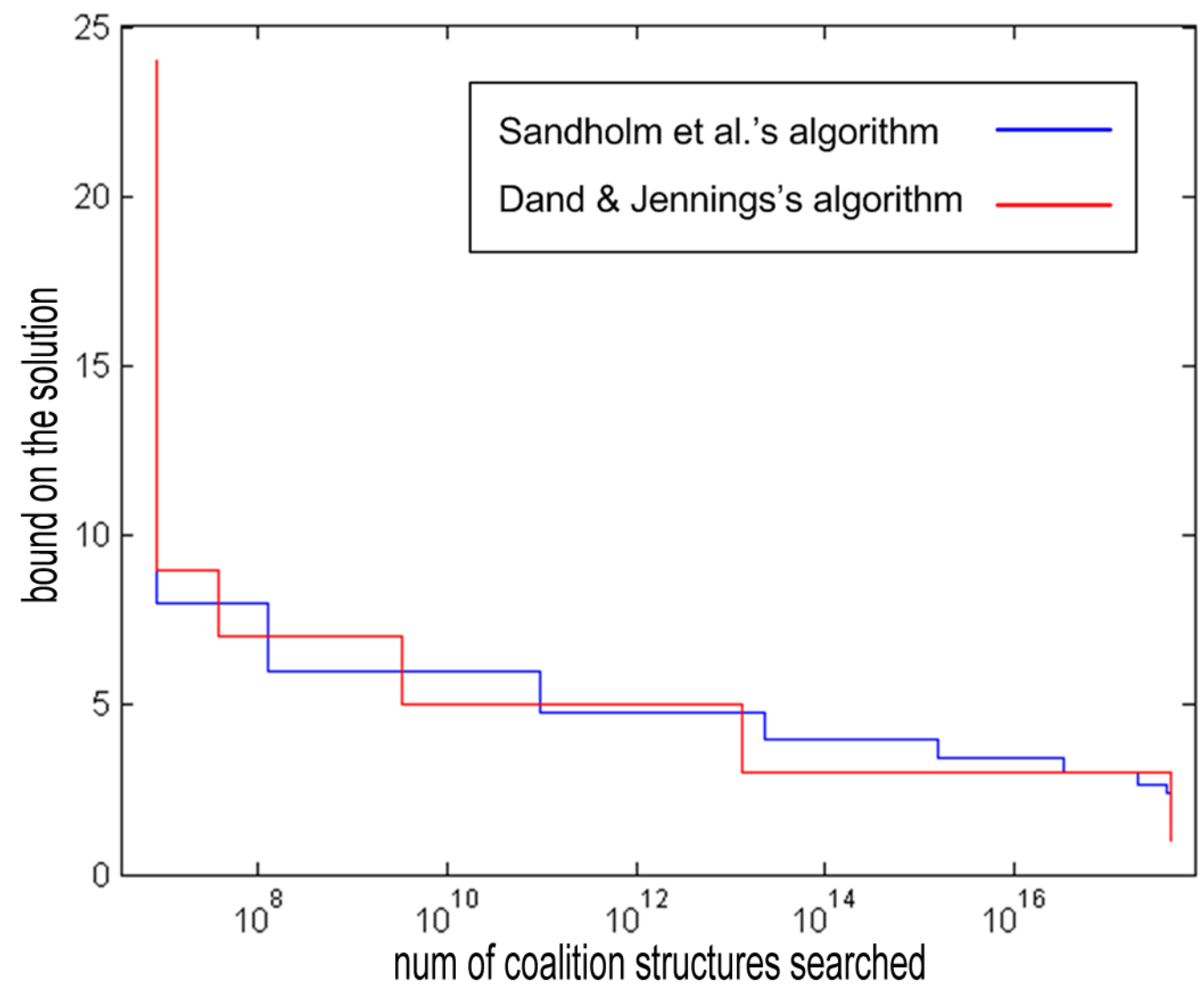

Figure 2: Given 24 agents, the figure shows, *on a log scale*, a comparison between the bound provided by Sandholm et al. (1999) and that provided by Dang and Jennings (2004), given different numbers of searched coalition structures.

a 0-1 integer programming problem), since any variable representing a possible coalition can either take a value of 1 (indicating that it belongs to the formed coalition structure) or 0 (indicating that it doesn't). Specifically, given $n$ agents, the *integer model* for the CSG problem can be formulated as follows:

$$\textit{Maximize} \sum_{i=1}^{2^n} v(C_i) \cdot x_i$$

$$\textit{subject to } Z \cdot X = e^T$$
$$X \in \{1, 0\}^n$$

where $Z$ is an $n \times 2^n$ matrix of zeros and ones, $X$ is a vector containing $2^n$ binary variables, and $e^T$ is the vector of $n$ ones. In more detail, every line in $Z$ represents an agent, and every column represents a possible coalition. As for $X$, having an element $x_i = 1$ corresponds to coalition $C_i$ being selected in the coalition structure. The first constraint ensures that the selected coalitions are both disjoint and exhaustive.

Such an integer programming problem is typically solved by applying *linear relaxation* coupled with *branch-and-bound* (Hillier & Lieberman, 2005). However, the main disadvantage of this approach is the huge memory requirement, which make it only applicable for small numbers of agents (see Section 5 for more details).

### 2.2 Non-Exact Algorithms for Coalition Structure Generation

These algorithms do not provide any guarantees on finding an optimal solution, nor do they provide worst-case guarantees on the quality of their solutions. Instead, they simply return *"good"* solutions. However, it is the fact that they can return a solution very quickly, compared to other algorithms, that often makes this class of algorithms more applicable, particularly given larger numbers of agents.

Generally speaking, as long as there is some regularity in the search space (i.e., the evaluation function is not arbitrary), genetic algorithms have the potential to detect that regularity and hence find the coalition structures that perform relatively effectively. To this end, Sen and Dutta (2000) developed a genetic algorithm for coalition structure generation. The algorithm starts with an initial set of candidate solutions (i.e. a set of coalition structures) called a population, which then gradually evolves towards better solutions. This is done in three main steps: evaluation, selection, and re-combination. In more detail, the algorithm evaluates every member of the current population, selects members based on their evaluation, and constructs new members from the selected ones by exchanging and modifying their contents. More details on the implementation can be found in the paper by Sen and Dutta (2000). The main advantage of this algorithm is that it can return solutions anytime, and that it scales up well with the increase in the number of agents. However, the main limitation is that the solutions it provides are not guaranteed to be optimal, or even guaranteed to be within a finite bound from the optimal. Moreover, even if the algorithm happens to find an optimal solution, it is not possible to verify this fact.

Another algorithm that belongs to this class of algorithms is the one developed by Shehory and Kraus (1998). This algorithm is greedy and operates in a decentralized manner. The heuristics they propose (in order to reduce the complexity of finding an optimal coalition structure) involve adding constraints on the size of the coalitions that are allowed to be formed. Specifically, only the coalitions up to a size $q < n$ are taken into consideration. The main advantage of this algorithm is that it can take into consideration overlapping coalitions.[5] Moreover, Shehory and Kraus prove that the solution they provide is guaranteed to be within a bound from the optimal solution. However, by optimal, they mean the best possible combination of all *permitted* coalitions. On the other hand, the algorithm provides no guarantees on the quality of its solutions compared to the actual optimal that could be found if *all* coalitions were taken into consideration.

To summarize, as discussed earlier, the main limitation of these algorithms is that they provide no guarantees on the solutions they generate while they search or when they terminate. However, these algorithms scale up well with the increase in the number of agents, making them particularly suitable for the cases where the number of agents is so large that no algorithm with exponential complexity can be executed in time.

After discussing the different approaches to the coalition structure generation problem, we can see that each of these approaches suffers from major limitations, making it either inefficient or inapplicable. This motivates our aim to develop more efficient CSG algorithms that can be applied to a wider range of problems, while taking into consideration the objectives outlined in Section 1. With this in mind, we first present in Section 3 a novel representation of the search space, and then present in Section 4 a novel algorithm that belongs to the first class of the aforementioned classification. As we will show, this algorithm avoids all the limitations that exist in state-of-the-art

5. A solution containing overlapping coalitions means that the agents may participate in more than one coalition at the same time.

algorithms belonging to this class, and meets all of the design objectives placed in Section 1 on CSG algorithms.

## 3. Search Space Representation

In this section, we describe our novel representation of the search space (i.e. the space of possible coalition structures). Recall that the space representation employed by most existing anytime algorithms is an undirected graph (see Figure 1 for an example), where the vertices represent coalition structures (Sandholm et al., 1999; Dang & Jennings, 2004). This representation, however, forces all possible solutions to be explored in order to guarantee that the optimal one has been found. Given this, we believe an ideal representation of the search space should allow the computation of solutions anytime, while establishing bounds on their quality, and should allow the pruning of the space to speed up the search. With this objective in mind, in this section we describe just such a representation. In particular, it supports an efficient search for the following reasons. First, it partitions the space into smaller, independent, sub-spaces for which we can identify upper and lower bounds, and thus, compute a bound on the solutions found during the search. Second, we can prune most of these sub-spaces since we can identify the ones that cannot contain a solution better than the best one found so far. Third, since the representation pre-determines the size of coalitions present in each sub-space, agents can balance their preference for certain coalition sizes against the cost of computing the solution for these sub-spaces. Next, we formally define our representation of the search space, describe its algebraic properties, and describe how to compute worst case bounds on the quality of the solution that our representation allows us to generate.

### 3.1 Partitioning the Search Space

We partition the search space $\mathcal{P}$ by defining sub-spaces that contain coalition structures that are similar according to some criterion. The particular criterion we specify here is based on the integer partitions of the number of agents.[6] Recall that an integer partition of $n$ is a multiset of positive integers that add up to exactly $n$ (Andrews & Eriksson, 2004). For example, given $n = 4$, the five distinct partitions are: $[4], [3, 1], [2, 2], [2, 1, 1]$, and $[1, 1, 1, 1]$.[7] It can easily be shown that the different ways to partition a set of $n$ elements can be directly mapped to the integer partitions of $n$, where the parts of the integer partition correspond to the cardinalities of the subsets (i.e. the sizes of the coalitions) within the set partition (i.e. coalition structure). For instance, the coalition structures $\{\{a_1, a_2\}, \{a_3\}, \{a_4\}\}$ and $\{\{a_4, a_1\}, \{a_2\}, \{a_3\}\}$ can be mapped to the integer partition $[2, 1, 1]$ since they each contain one coalition of size 2, and two coalitions of size 1. We define the aforementioned mapping by the function $F : \mathcal{P} \to \mathcal{G}$, where $\mathcal{G}$ is the set of integer partitions of $n$. Thus, $F$ defines an equivalence relation $\sim$ on $\mathcal{P}$ such that $CS \sim CS''$ iff $F(CS) = F(CS'')$ (i.e. the sizes of the coalitions in $CS$ are the same as those in $CS''$). Given this, the pre-image[8] of an integer partition $G$, noted as $P_G = F^{-1}[\{G\}]$, contains all the coalition structures that correspond

6. Other criteria could be developed to further partition the space into smaller sub-spaces, but the one we develop here allows us to choose coalition structures with certain properties as we show later.

7. For presentation clarity, square brackets are used throughout the paper (instead of the curly ones) to distinguish between multisets and sets.

8. Recall that the pre-image or inverse image of $G \subseteq \mathcal{G}$ under $F : \mathcal{P} \to \mathcal{G}$ is the subset of $\mathcal{P}$ defined by $F^{-1}[\{G\}] = \{CS \in \mathcal{P} | F(CS) = G\}$.

to the same integer partition $G$. Every such pre-image represents a sub-space in our representation. This implies that the number of sub-spaces in our representation is the same as the number of possible integer partitions, which grows exponentially with $n$. This number, however, remains insignificant compared to the number of possible coalitions and coalition structures (e.g., given 24 agents, the number of possible integer partitions is only 1575, while the number of possible coalitions is $16777215$, and the number of possible coalition structures is nearly $4.4 \times 10^{17}$).

We categorize the sub-spaces into levels based on the number of parts within the integer partitions. Specifically, level $\mathcal{P}_i = \{P_G : |G| = i\}$ contains all the sub-spaces that correspond to an integer partition with $i$ parts (see Figure 3 for an example of 4 agents).[9] In what follows, we show how to compute bounds for the sub-spaces ($P_G : G \in \mathcal{G}$) in our representation.

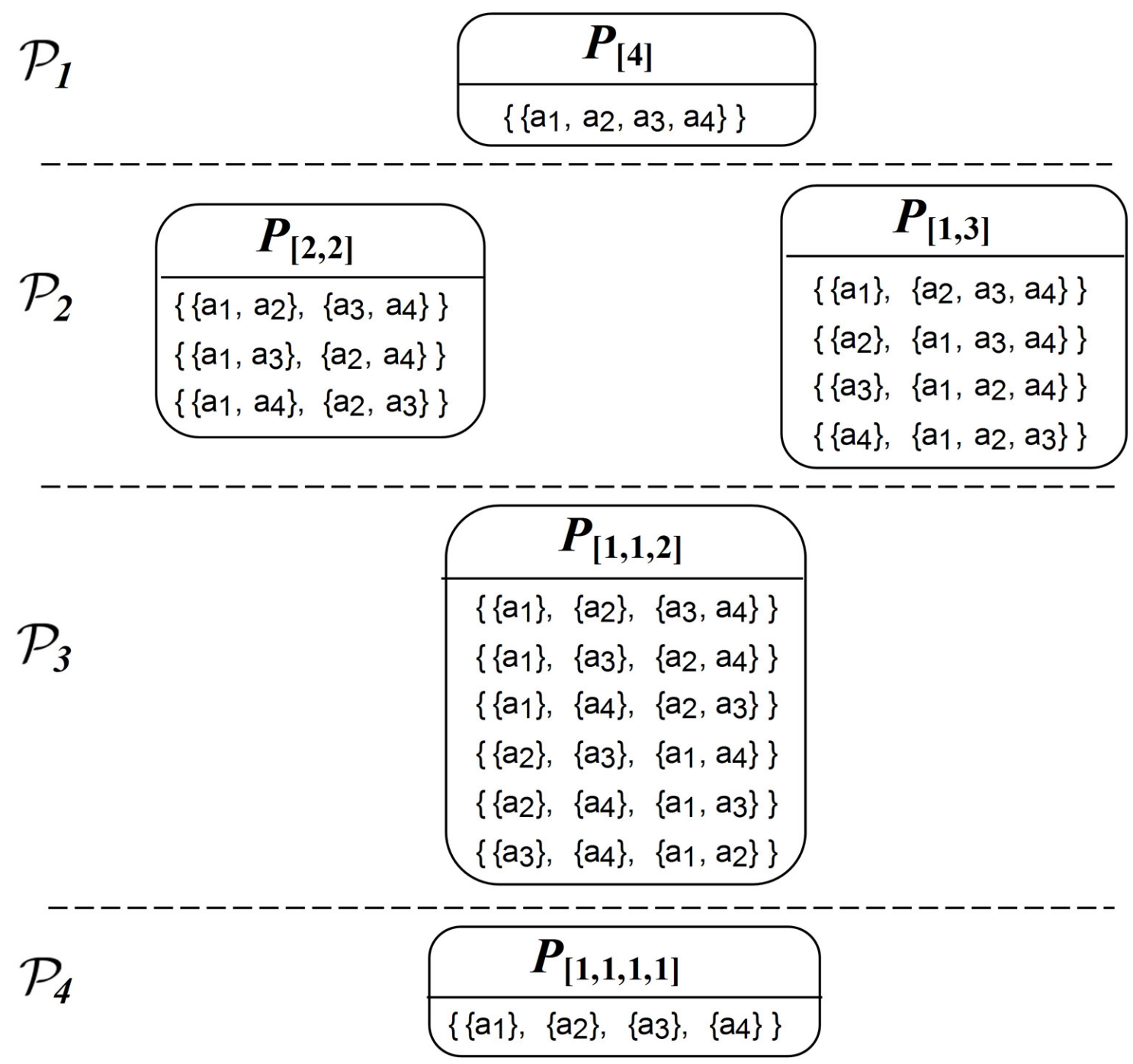

Figure 3: An example of our representation of the search space given 4 agents.

### 3.2 Computing Bounds for Sub-spaces

For each sub-space $P_G$, it is possible to compute an upper and a lower bound on the value of the best[10] coalition structure that could be found in it. To this end, let $L_s$ be the list of coalitions of size $s$, and let $max_s, min_s$, and $avg_s$, be the maximum, minimum, and average value of the coalitions in $L_s$ respectively. Moreover, given an integer partition $G$, let $T_G$ be the Cartesian product of the lists $L_s : s \in G$. That is, $T_G = \prod_{s \in G} (L_s)^{G(s)}$, where $G(s)$ is the multiplicity of $s$ in

9. Note that the levels in our representation are basically the same as those that appear in the coalition structure graph, except that the coalition structures within each level are now categorized into sub-spaces. In other words, the coalition structures that belong to the sub-spaces in $\mathcal{P}_i$ are the same as those that belong to $LV_i$.

10. Throughout this paper, a coalition structure is described as being *"the best"* if it has the highest value.

$G$. For example, given $G = [5,4,4,4,1,1]$, we have $T_G = (L_5)^1 \times (L_4)^3 \times (L_1)^2$. Note that $T_G$ contains many combinations of coalitions that are considered invalid coalition structures. This is because some of the coalitions within these combinations may overlap. For example, $T_{[2,1,1]}$ contains the following combination, $\{\{a_1, a_2\}, \{a_1\},\{a_3\}\}$, which is not a valid coalition structure because agent $a_1$ appears in two coalitions. Now, if we consider the value of each element (i.e. combination of coalitions) in $T_G$ to be the sum of the values of all the coalitions in that element, then the maximum value that an element in $T_G$ can take, denoted $MAX_G$, is computed as follows: $MAX_G = \sum_{s \in G} max_s \times G(s)$. Based on this, it is easy to demonstrate that $MAX_G$ is an upper bound on the value of the best coalition structure in $P_G$ (since $P_G$ is a subset of $T_G$).

Similarly, the minimum value that an element in $T_G$ can take, denoted $MIN_G$, is computed as follows: $MIN_G = \sum_{s \in G} min_s \times G(s)$. Although this could intuitively be considered a lower bound on the value of the best coalition structure (i.e. solution) in $P_G$, we show that it is actually possible to compute a higher (i.e. better) lower bound than $MIN_G$.

In more detail, let $AVG_G$ be the average value of all the coalition structures in $P_G$. Then, $AVG_G$ would be a lower bound on the value of the best coalition structure in $P_G$ (since an average is always greater than, or equal to, a minimum). The key point to note, here, is that we can compute $AVG_G$ without having to go through *any* of the coalition structures in $P_G$. Instead, we can compute it by simply summing the averages of the coalition lists (see Theorem 1), and these averages can be computed immediately after scanning the input, which is significantly smaller than the space of possible coalition structures.

**Theorem 1.** *Let $G = [g_1, \cdots, g_i, \cdots, g_{|G|}]$ be an integer partition, and let $AVG_G$ be the average of the values of all the coalition structures in $P_G$. Also, let $avg_{g_i}$ be the average of the values of all the coalitions in $L_{g_i}$. Then, the following holds:*

$$AVG_G = \sum_{i=1}^{|G|} avg_{g_i}$$

**Proof.** See Appendix B.

Having described our novel representation of the search space, we present (in the following section) an anytime algorithm that uses this representation to search through the possible coalitions structures to eventually find an optimal one.

## 4. Solving the Coalition Structure Generation Problem

Assuming that the value of every coalition $C$ is given by a characteristic function $v(C) \in \mathbb{R}$, and that the value of every coalition structure is given by the function $V(CS) = \sum_{C \in CS} v(C)$, our goal is to search through the set of possible coalition structures, noted as $\mathcal{P}$, in order to find an optimal coalition structure which is computed as:

$$CS^* = \arg\max_{CS \in \mathcal{P}} V(CS) \tag{1}$$

given $v(C)$ for all $C \in 2^A \backslash \{\emptyset\}$. Note that, in this section, the terms *"coalition structure"* and *"solution"* will be used interchangeably.

Basically, our novel anytime Integer-Partition based algorithm (which we call IP) consists of the following two main steps:

1. Scanning the input in order to compute the bounds (i.e. $MAX_G$ and $AVG_G$) for every sub-space $P_G$ — while doing so, we can (at a very small cost):
   (a) find the best coalition structures within particular sub-spaces.
   (b) prune other sub-spaces based on their upper-bounds.
   (c) establish a worst-case bound on the quality of the best solution found so far.
2. Searching within the remaining sub-spaces — the techniques we use allow us to:
   (a) avoid making unnecessary comparisons between coalitions to generate valid coalition structures (i.e. those that contain disjoint coalitions).
   (b) avoid computing the same coalition structure more than once.
   (c) apply branch-and-bound to further reduce the amount of search to be done.

The following sub-sections describe each of the aforementioned steps in more detail. To this end, we will use $CS'$ to denote the best coalition structure found so far, and $\mathcal{G}' \subseteq \mathcal{G}$ to denote the integer partitions that represent the sub-spaces that have not been searched.

### 4.1 Scanning the Input

The input to the coalition structure generation problem is the value associated to each coalition, i.e. $v(C)$ for all $C \in 2^A \backslash \{\emptyset\}$. One way of representing this input is to use a table containing every coalition along with its value. Another way is to agree on an ordering of the coalitions, and to use a list containing only the values of these ordered coalitions (i.e. the first value in the list corresponds to the first coalition, the second value corresponds to the second coalition, and so on). We use the latter representation since it does not require maintaining the coalitions themselves in memory. In more detail, we assume that the input is given as follows: $\mathbf{v}(L_s)\ \forall s \in \{1, 2, \ldots, n\}$, where $\mathbf{v}(L_s)$ is a list containing the values of all the coalitions of size $s$. Moreover, we assume that the coalitions in $L_s$ are ordered lexicographically. For example, coalition $\{a_1, a_2, a_4\}$ has its elements ordered according to their indices, and the coalition itself is found above $\{a_1, a_2, a_3\}$ and below $\{a_1, a_3, a_4\}$ in the list $L_3$ (this is depicted in Figure 4). This ordering can easily be generated using the techniques that are used by Rahwan and Jennings (2007). Next, we describe the individual steps of the algorithm that depicts the scanning process (see Algorithm 1).

At first, we scan the value of the one coalition of size $n$ (i.e. the grand coalition). This would be the value of the only coalition structure in $P_{[n]}$ (which is the only sub-space in $\mathcal{P}_1$). After that, we scan the values of all the coalitions of size 1 (i.e. singleton coalitions), and by summing these values, we get the value of the only coalition structure in $P_{[1,1,\ldots,1]}$ (which is the only sub-space in $\mathcal{P}_n$). At this point (step 1), it is possible to compute the best coalition structure found so far (i.e. $CS'$).

Having searched through levels $\mathcal{P}_1$ and $\mathcal{P}_n$, we now show how to search through level $\mathcal{P}_2$ at a very low cost while scanning the input. To this end, let $\mathcal{G}^2 = \{G \in \mathcal{G} : |G| = 2\}$ be the set of integer partitions that contain two parts each. Then, as a result of the assumed ordering of the input, any two complementary coalitions $C$ and $\widehat{C}$ in a coalition structure $CS = \{C, \widehat{C}\}$ are always diametrically positioned in the coalition lists $L_{|C|}$ and $L_{|\widehat{C}|}$, and that happens even if $|C| = |C'|$. For example, given 6 agents, the coalitions $\{a_1\}$ and $\{a_2, a_3, a_4, a_5, a_6\}$ are diametrically positioned in

**Algorithm 1 :** `scanAndSearch()` – scan the input, generate initial solutions and bounds.

**Require:** $n,\ \{\mathbf{v}(L_s)\}_{s\in\{1,2,...,n\}}$

$CS' \leftarrow \arg\max_{CS\in\{\ \{a_1,...,a_n\},\ \{\{a_1\},...,\{a_n\}\}\ \}} V(CS)$

**for** $s = 1$ to $\lfloor \frac{n}{2} \rfloor$ **do**

  $\widehat{s} \leftarrow n - s$

  **if** $s = \widehat{s}$ `{if cycling through the same list.}` **then**

    $end = \lfloor |\mathbf{v}(L_s)|/2 \rfloor$

  **else**

    $end = |\mathbf{v}(L_s)|$

  **end if**

  Set $max_s,\ max_{\widehat{s}},\ v_{max}$ to $-\infty$ , and set $sum_s,\ sum_{\widehat{s}}$ to 0

  **for** $x = 1$ to $end$ `{cycle through the lists` $\mathbf{v}(L_s)$ `and` $\mathbf{v}(L_{\widehat{s}})$`.}` **do**

    $\widehat{x} \leftarrow |\mathbf{v}(L_s)| - x + 1$

    $v \leftarrow \mathbf{v}(L_s)^x$ , $\widehat{v} \leftarrow \mathbf{v}(L_{\widehat{s}})^{\widehat{x}}$ `{extract element at` $x$,$\widehat{x}$ `from` $\mathbf{v}(L_s), \mathbf{v}(L_{\widehat{s}})$`.}`

    **if** $v_{max} < v + \widehat{v}$ **then**

      $v_{max} \leftarrow v + \widehat{v}$

      $x_{max} = x$ `{record the index in` $\mathbf{v}(L_s)$ `at which` $v$ `is located.}`

    **end if**

    **if** $max_s < v$ **then**

      $max_s \leftarrow v$ `{record the maximum value in` $\mathbf{v}(L_s)$`.}`

    **end if**

    **if** $max_{\widehat{s}} < \widehat{v}$ **then**

      $max_{\widehat{s}} \leftarrow \widehat{v}$ `{record the maximum value in` $\mathbf{v}(L_{\widehat{s}})$`.}`

    **end if**

    $sum_s \leftarrow sum_s + v$ , $sum_{\widehat{s}} \leftarrow sum_{\widehat{s}} + \widehat{v}$

  **end for**

  $\widehat{x}_{max} \leftarrow |\mathbf{v}(L_s)| - x_{max} + 1$

  **if** $V(CS') < V(\{L_s^{x_{max}}, L_{\widehat{s}}^{\widehat{x}_{max}}\})$ **then**

    $CS' \leftarrow \{L_s^{x_{max}}, L_{\widehat{s}}^{\widehat{x}_{max}}\}$ `{update the best coalition structure found so far.}`

  **end if**

  $avg_s \leftarrow sum_s/|\mathbf{v}(L_s)|$ , $avg_{\widehat{s}} \leftarrow sum_{\widehat{s}}/|\mathbf{v}(L_{\widehat{s}})|$ `{compute averages.}`

**end for**

$\mathcal{G}' \leftarrow \mathcal{G} \setminus \mathcal{G}^2$

**for** $G \in \mathcal{G}'$ `{compute upper and lower bounds for each sub-space in` $\mathcal{G}'$`.}` **do**

  $MAX_G \leftarrow \sum_{s\in G} max_s \cdot G(s)$

  $AVG_G \leftarrow \sum_{s\in G} avg_s \cdot G(s)$

**end for**

$UB^* \leftarrow \max[\ V(CS'),\ \max_{G\in\mathcal{G}'}[MAX_G]\ ]$

$LB^* \leftarrow \max[\ V(CS'),\ \max_{G\in\mathcal{G}'}[AVG_G]\ ]$

$\mathcal{G}' \leftarrow$ `prune`$(\mathcal{G}', \{MAX_G\}_{G\in\mathcal{G}'}, LB^*)$ `{prune the sub-spaces that have an upper bound lower than` $LB^*$`.}`

$\beta \leftarrow \min[\ n/2\ ,\ UB^*/V(CS')\ ]$ `{compute a worst-case bound on` $V(CS')$`.}`

**return** $CS', \beta, \{max_s\}_{s\in\{1,...,n\}}, \mathcal{G}', \{MAX_G\}_{G\in\mathcal{G}'}, \{AVG_G\}_{G\in\mathcal{G}'}$

the lists $L_1$ and $L_5$ respectively, and the coalitions $\{a_1, a_2, a_3\}$ and $\{a_4, a_5, a_6\}$ are diametrically positioned in the list $L_3$ (see Figure 4 for an example of 6 agents).

| $L_1$ | $L_2$ | $L_3$ | $L_4$ | $L_5$ | $L_6$ |
|---|---|---|---|---|---|
| $a_1$ | $a_1, a_2$ | $a_1, a_2, a_3$ | $a_1, a_2, a_3, a_4$ | $a_1, a_2, a_3, a_4, a_5$ | $a_1, a_2, a_3, a_4, a_5\ a_6$ |
| $a_2$ | $a_1, a_3$ | $a_1, a_2, a_4$ | $a_1, a_2, a_3, a_5$ | $a_1, a_2, a_3, a_4, a_6$ | |
| $a_3$ | $a_1, a_4$ | $a_1, a_2, a_5$ | $a_1, a_2, a_3, a_6$ | $a_1, a_2, a_3, a_5, a_6$ | |
| $a_4$ | $a_1, a_5$ | $a_1, a_2, a_6$ | $a_1, a_2, a_4, a_5$ | $a_1, a_2, a_4, a_5, a_6$ | |
| $a_5$ | $a_1, a_6$ | $a_1, a_3, a_4$ | $a_1, a_2, a_4, a_6$ | $a_1, a_3, a_4, a_5, a_6$ | |
| $a_6$ | $a_2, a_3$ | $a_1, a_3, a_5$ | $a_1, a_2, a_5, a_6$ | $a_2, a_3, a_4, a_5, a_6$ | |
| | $a_2, a_4$ | $a_1, a_3, a_6$ | $a_1, a_3, a_4, a_5$ | | |
| | $a_2, a_5$ | $a_1, a_4, a_5$ | $a_1, a_3, a_4, a_6$ | | |
| | $a_2, a_6$ | $a_1, a_4, a_6$ | $a_1, a_3, a_5, a_6$ | | |
| | $a_3, a_4$ | $a_1, a_5, a_6$ | $a_1, a_4, a_5, a_6$ | | |
| | $a_3, a_5$ | $a_2, a_3, a_4$ | $a_2, a_3, a_4, a_5$ | | |
| | $a_3, a_6$ | $a_2, a_3, a_5$ | $a_2, a_3, a_4, a_6$ | | |
| | $a_4, a_5$ | $a_2, a_3, a_6$ | $a_2, a_3, a_5, a_6$ | | |
| | $a_4, a_6$ | $a_2, a_4, a_5$ | $a_2, a_4, a_5, a_6$ | | |
| | $a_5, a_6$ | $a_2, a_4, a_6$ | $a_3, a_4, a_5, a_6$ | | |
| | | $a_2, a_5, a_6$ | | | |
| | | $a_3, a_4, a_5$ | | | |
| | | $a_3, a_4, a_6$ | | | |
| | | $a_3, a_5, a_6$ | | | |
| | | $a_4, a_5, a_6$ | | | |

Figure 4: An example of the assumed ordering of the coalition lists.

Based on this, for every integer partition $G = [g_1, g_2] \in \mathcal{G}^2$, we compute the values of all the coalition structures in $P_G$ by simply summing the values of the coalitions as we scan the lists $\mathbf{v}(L_{g_1})$ and $\mathbf{v}(L_{g_2})$, starting at different extremities for each list. Once these lists have been scanned (steps 10 to 24), it is possible to obtain the two values of which the sum is maximized. Moreover, it is possible to obtain the indices in the lists at which these values are located (see how $x_{max}$ and $\widehat{x}_{max}$ are computed in steps 15 and 25 respectively). Then, by obtaining these indices, we know where in $L_{g_1}$ and $L_{g_2}$ to find the two coalitions that belong to the best coalition structure in $P_{[g_1,g_2]}$ (this comes from the fact that the position of any value in $\mathbf{v}(L_s) : s \in \{1, ..., n\}$ is exactly the position of the corresponding coalition in $L_s$).

Note, however, that the input includes only $\mathbf{v}(L_{g_1})$ and $\mathbf{v}(L_{g_2})$ (i.e. it does not include $L_{g_1}$ and $L_{g_2}$). For this reason, an algorithm is required that can return a coalition $C$ given its position in the ordered list $L_{|C|}$. Rahwan and Jennings (2007) have developed a polynomial-time algorithm that does exactly that. Therefore, we use it to find the required coalitions and compose the best coalition structure in $P_{\{g_1,g_2\}}$ (see steps 26 and 27).[11]

While scanning $\mathbf{v}(L_{g_1})$ and $\mathbf{v}(L_{g_2})$, we also compute $max_{g_1}$ and $max_{g_2}$ (steps 17 to 22), as well as $avg_{g_1}$ and $avg_{g_2}$ (step 29). Note that, in Algorithm 1, we scan $\mathbf{v}(L_s)$ and $\mathbf{v}(L_{n-s})$ for all $s \in \{1, \dots, \lfloor \frac{n}{2} \rfloor\}$ and this implies that $max_s$ and $avg_s$ are computed for all $s \in \{1, \dots, n\}$. Also note that this whole process is linear in the size of the input (i.e. $O(y)$ where $y = 2^n - 1$ is the size of the input).

11. By $L_s^x$ we mean that we extract the element at position $x$ from $L_s$.

Having computed $max_s$ and $avg_s$ for every size $s$, we can now compute upper and lower bounds for every sub-space (as in steps 32 to 34). By using these bounds, it is possible to compute an upper bound $UB^*$ and a lower bound $LB^*$ on the value of the optimal coalition structure (see steps 36 and 37). Hence, every sub-space $P_G$ that has an upper bound $MAX_G < LB^*$ can be pruned straight away. The prune function (used in step 38) is implemented as in Algorithm 2.

**Algorithm 2 :**`prune` $(\mathcal{G}', \{MAX_G\}_{G \in \mathcal{G}'}, v)$ – prune sub-spaces.

```
for G ∈ G' do
  if MAX_G ≤ v {if the upper bound of P_G is lower than v.} then
    G' ← G' \ G  {remove G.}
  end if
end for
return G'
```

Another advantage of our scanning procedure is that it allows us to compute a worst-case bound $\beta$ on the value of $CS'$ as follows: $\beta = min(\ \frac{n}{2},\ \frac{UB^*}{V(CS')}\ )$ (see step 39). This comes from the fact Sandholm et al. (1999) have proved that the value of the best coalition structure in levels $LV_1$, $LV_2$ and $LV_n$ (corresponding to $\mathcal{P}_1, \mathcal{P}_2$, and $\mathcal{P}_n$ respectively) is within a bound $\frac{n}{2}$ from the optimal.

So far, by only scanning the input, we have calculated $max_s$ and $avg_s$ for all $s \in \{1, \ldots, n\}$, we have searched levels $\mathcal{P}_1, \mathcal{P}_2, \mathcal{P}_n$, we have calculated $MAX_G$ and $AVG_G$ for all the sub-spaces within the remaining levels (i.e. $\mathcal{P}_3, ..., \mathcal{P}_{n-1}$), we have pruned some of these sub-spaces, and we have established a worst-case bound $\beta$ on the quality of the best solution found so far. Moreover, it is possible to specify a bound $\beta^* \geq 1$ within which any solution is acceptable. In more detail, if the best solution found so far fits within the specified bound (i.e. if $\beta \leq \beta^*$) then no further search is required. Otherwise, the sub-spaces that have not been pruned (if there are any) must be searched. Next, we specify how this search is done.

### 4.2 Selecting and Searching a Sub-Space

Given the set of sub-spaces left after scanning the input, we select a sub-space to be searched, and we find the best coalition structure in it. After that, we prune all the remaining sub-spaces that have an upper bound lower than the best value found so far. This process of selecting, searching, and pruning, is repeated until either of the following termination conditions is reached:

- The best coalition structure found so far fits within the specified bound $\beta^*$.
- All the remaining sub-spaces have either been searched or pruned.

This can be seen in Algorithm 3. Basically, the algorithm works as follows. A sub-space $P_{G''}$ is selected to be searched (step 2).[12] Once $P_{G''}$ has been searched (step 3), it is removed from the set of remaining sub-spaces (step 4). After that, we check whether $CS'$ has been modified during the search (step 5), and, if that is the case, then every sub-space with an upper bound lower than $V(CS')$ is pruned (step 6).[13] $UB^*$ and $\beta$ are then updated in steps 8 and 9 respectively, and if the current

12. In step 2, we actually select an integer partition, but this implies that the corresponding sub-section is to be searched.

13. Checking whether $CS'$ belongs to $P_{G''}$ can easily be done by checking whether the sizes of the coalitions in $CS'$ match the parts in $G''$.

best solution fits within the specified bound $\beta^*$ then it is returned (steps 10 and 11). Otherwise, this whole process is repeated given the remaining sub-spaces (if there are any). In what follows, we further elaborate on the sub-space selection strategy and the sub-space search algorithm since these are the key parts of this algorithm.

**Algorithm 3 :**`searchSpace()` – search, or prune, the remaining sub-spaces.

**Require:** $\mathcal{G}', \{MAX_G\}_{G \in \mathcal{G}'}, A, \beta^*$

1: **while** $\mathcal{G}' \neq \emptyset$ **do**
2:   **Select** $G''$ {`select the integer partition that represents the next sub-space to be searched.`}
3:   $CS' \leftarrow \texttt{searchList}(G'', 1, 1, A, CS', \overrightarrow{CS})$ {`search within` $P_{G''}$ `and update` $CS'$.}
4:   $\mathcal{G}' \leftarrow \mathcal{G}' \setminus G''$ {`remove` $P_{G''}$ `from the list of sub-spaces that are yet to be searched.`}
5:   **if** $CS' \in P_{G''}$ {`If` $CS'$ `has been modified while searching` $P_{G''}$.} **then**
6:     $\mathcal{G}' \leftarrow \texttt{prune}(\mathcal{G}', \{MAX_G\}_{G \in \mathcal{G}'}, V(CS'))$ {`prune the sub-spaces that have upper bounds lower than` $V(CS')$.}
7:   **end if**
8:   $UB^* \leftarrow \max[\, V(CS'),\ \max_{G \in \mathcal{G}'}[MAX_G]\,]$ {`update the upper bound on the value of the optimal coalition structure(s).`}
9:   $\beta \leftarrow \min[\frac{UB^*}{V(CS')}, \beta]$ {`update the worst-case bound on` $V(CS')$.}
10:   **if** $\beta \leq \beta^*$ {`if` $CS'$ `is within the specified bound from the optimal.`} **then**
11:     **return** $CS'$
12:   **end if**
13: **end while**
14: **return** $CS'$

#### 4.2.1 SELECTING A SUB-SPACE

It can easily be seen that, unless we search the sub-spaces that have an upper bound greater than $V(CS')$, we cannot verify that $CS'$ is an optimal solution. This implies that $\beta$ remains greater than 1 until the following sub-spaces are searched: $\{P_G : MAX_G \geq V(CS^*)\}$. This can be done by selecting the next sub-space to be searched using the following selection rule:

$$\textbf{Select } \; G = \arg\max_{G \in \mathcal{G}'}(MAX_G)$$

As a result of this selection strategy, all sub-spaces with an upper bound lower than $V(CS^*)$ will not be searched and these can constitute a significant portion of the search space (see Section 5.3 for more details). Another result is that it will always be beneficial to search a sub-space, even if that sub-space does not contain a better solution than the one found so far. This is because the above selection strategy ensures that $UB^*$ is reduced whenever a sub-space is searched, and this improves the worst-case guarantee $\beta$ on the quality of the current best solution.

Note that this selection rule is mainly for the cases where an *optimal solution* is sought. In case we are after a near-optimal solution where a bound $\beta^* > 1$ is specified (e.g., $\beta^* = 1.05$ means that the solution sought needs to have a value that is at least $95\%$ of the optimal one), then other selection rules may be used. For example, one could select to search the smallest sub-space that could, potentially, give a value greater than or equal to $\frac{UB^*}{\beta^*}$ (hoping to find an acceptable solution in the least amount of search). This can be expressed as:

$$\textbf{Select } G = \underset{G \in \mathcal{G}' : UB_G \geq \frac{UB^*}{\beta^*}}{\arg\min} (|P_G|)$$

where $|P_G|$ is the size of (i.e. the number of coalition structures in) $P_G$. More specifically, $|P_G|$ is computed as follows:

**Theorem 2.** *Let $G = [g_1, \ldots, g_{|G|}]$ be an integer partition, and let $|P_G|$ be the number of coalition structures in $P_G$. Moreover, let $C^n_s$ be the binomial coefficient,*[14] *and let $E(G)$ be the underlying set of $G$.*[15] *Then, the following holds:*

$$|P_G| = \frac{C^n_{g_1} \times C^{n-g_1}_{g_2} \times \ldots \times C^{n-(g_1+\ldots+g_{|G|-1})}_{g_{|G|}}}{\prod_{s \in E(G)} G(s)!}$$

**Proof.** See Appendix C.

The key point to note is that, given our representation, we can specify $\beta^*$ in cases where computing the optimal solution would be too costly and, given this, we can modify the selection rule accordingly to speed up the search.

Another advantage of being able to control the sub-spaces to be searched is that the agents can choose what types of coalition structures to build according to their computational resources or private preferences. For example, it has been argued that the computation time could be reduced if we limit the size of the coalitions that can be formed (Shehory & Kraus, 1998). However, this is a very costly, self-imposed constraint since it possibly means neglecting a number of highly efficient solutions. Instead, by using IP, it is possible to determine, *ex-ante* (i.e. before performing the search), which sub-spaces are most promising according to their upper and lower bounds. Therefore the computation time can be focused on these sub-spaces and the gains can be traded-off against the computation time.

In some other cases, agents may need to form $q$ coalitions (Shehory & Kraus, 1995). For example, they may need to perform $q$ tasks and therefore need to divide up into $q$ teams to perform these tasks separately. Moreover, they may wish to have coalitions with a maximum size of $z$ as they may have certain constraints on the amount of resources available to each coalition. By using our representation, such preferences can be naturally expressed and the search can be directed to fit these preferences transparently. Formally, our search space can easily be redefined as follows:

$$\mathcal{G}'' = \{G \in \mathcal{G} : |G| = m \ \wedge \ \forall g \in G : |g| \leq z\}$$

In all the above cases where agents can express preferences for coalition structures of certain sizes, they can now, *a priori*, balance such preferences with the quality of the solutions that can be

14. Recall that the binomial coefficient represents the number of possible combinations of size $s$ taken from $n$ elements, and is computed as follows: $C^n_s = \frac{n!}{k!(n-k)!}$, where $n!$ is the factorial of $n$.

15. Recall that the underlying set $E(G)$ of a multiset $G$ is a subset of $G$ in which each element in $G$ appears only once in $E(G)$. For example, $\{1, 2\}$ is the underlying set of $[1, 1, 2]$.

obtained. This is because we are able to determine the worst-case bound from the optimal that the search of a given sub-space will generate (i.e. $\frac{UB^*}{AVG_G}$). We next describe how we search through the chosen sub-space.

#### 4.2.2 SEARCHING A SUB-SPACE

Given an integer partition $G = [g_1, g_2, \cdots, g_{|G|}] \in \mathcal{G}$, we need to cycle through the coalition structures that belong to $P_G$ in order to find the best one. Here, without loss of generality, we assume that $g_1 \leq g_2 \leq \cdots, \leq g_{|G|}$. Perhaps the most obvious way of performing this cyclation process is shown in Figure 5. Here, a variable $\overrightarrow{CS} = \langle C_1, C_2, \cdots, C_{|G|} \rangle$ is used to cycle though the coalition structures in $P_G$ as follows. First, $C_1$ is assigned to one of the coalitions in $L_{g_1}$. After that, $C_2$ is used to cycle through $L_{g_2}$ until a coalition that does not overlap with $C_1$ is found. After that, $C_3$ is used to cycle through $L_{g_3}$ until a coalition that does not overlap with $\{C_1, C_2\}$ is found. This is repeated until every $C_k \in \overrightarrow{CS}$ is assigned to a coalition in $L_{g_k}$. In this case, $\overrightarrow{CS}$ would be a valid coalition structure belonging to $P_G$. The value of this coalition structure is then calculated and compared with the maximum value found so far. After that, the coalitions in $\overrightarrow{CS}$ are updated so as to set $\overrightarrow{CS}$ to another coalition structure in $P_G$. Here, a coalition $C_k$ is only updated once we have examined all the possible instances of $C_{k+1}, \ldots, C_{|G|}$ that do not overlap with $\{C_1, \ldots, C_k\}$. For example, in Figure 5, we only update $C_2$ (step 5 in the figure) once we have examined all the possible instances of $C_3$ that do not overlap with $\{C_1, C_2\}$ (steps 2, 3, 4 in the figure). This ensures that $\overrightarrow{CS}$ is assigned to different coalition structures, and that, eventually, every possible coalition structure in $P_G$ is examined.

Next, we show how this process can be done without storing any of the lists $L_{g_1}, L_{g_2}, \cdots, L_{g_{|G|}}$ in memory. To this end, let $LC^n_{g_k} \;:\; 1 \leq g_k \leq n$ be the list of combinations of size $g_k$ that are taken from the set $\{1, 2, \cdots, n\}$, where the combinations are ordered lexicographically in the list. Given this, both $LC^n_{g_k}$ and $L_{g_k}$ contain the subsets of size $g_k$ that are taken from a set of size $n$. The only difference is that $LC^n_{g_k}$ is a list of combinations of numbers while $L_{g_k}$ is a list of coalitions of agents. Now, Rahwan and Jennings (2007) have shown how to cycle through the combinations in $LC^n_{g_k}$ without storing the entire list in memory. Instead, only one combination is stored at a time. This is based on the assumed ordering which implies that the last combination in $LC^n_{g_k}$ is always: $\{1, 2, \cdots, g_k\}$. This ordering also implies that, given any combination located at index $x$ in the list, where $1 < x \leq \left|LC^n_{g_k}\right|$, it is possible to compute the combination located at index $x - 1$ (for more details, see the paper by Rahwan & Jennings, 2007). Hence, in order to go through the coalitions in $L_{g_k}$, we use a variable $M_k$ to cycle[16] through the combinations in $LC^n_{g_k}$ and, for every instance of $M_k$, we extract the corresponding coalition $C_k \in L_{g_k}$ using the following operation:

$$C_k = \{a_i \in A \mid i \in M_k\} \tag{2}$$

For example, given that $M_k = \{2, 4, 5\}$, the corresponding coalition would be $\{a_2, a_4, a_5\}$. Since there is a direct mapping (as defined by equation 2) from every combination in $LC^n_{g_k}$ to a coalition in $L_{g_k}$, then, by having $M_k$ cycle through every combination in $LC^n_{g_k}$, we cycle through all the coalitions in $L_{g_k}$.

16. This can be done by initializing $M_k$ to the last combination in $LC^n_{g_k}$ (i.e. to $\{1, 2, \cdots, g_k\}$), and then iteratively shifting $M_k$ up in the list as in the paper by Rahwan and Jennings (2007), until every combination in $LC^n_{g_k}$ is examined.

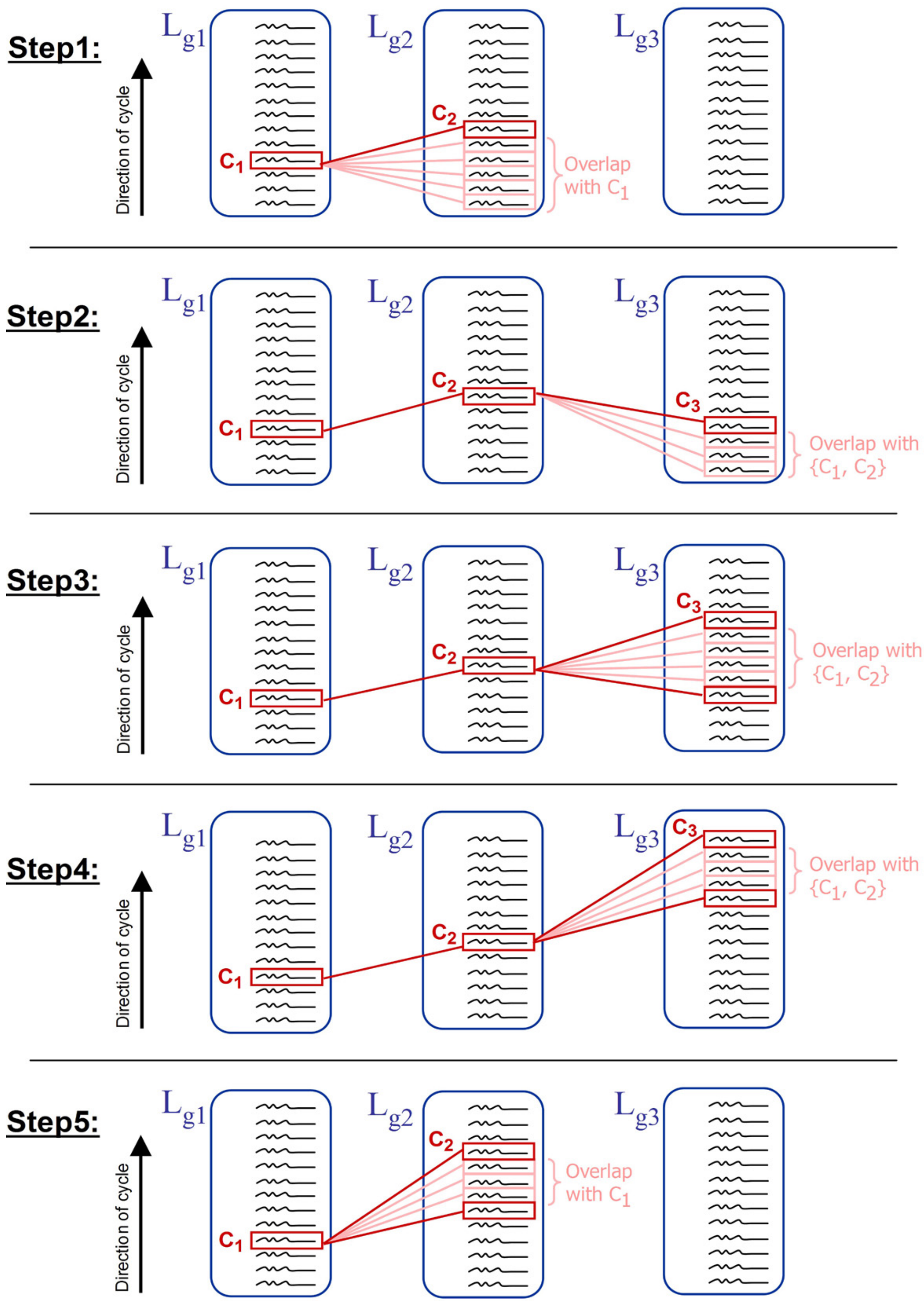

Figure 5: A naïve cyclation process for cycling through the coalition structures in a sub-space.

Intuitively, this *naïve cyclation process*, which we call NCP, can be viewed as being efficient. After all, what we need is to find the coalition structure in $P_G$ that has the maximum value, and NCP guarantees to find such a coalition structure. However, it suffers from the following major limitations:

1. NCP works by searching through the ordered sets in $T_G$ — the Cartesian product of the lists $L_s : s \in G$ — in order to find those that belong to $P_G$ (i.e. those that contain disjoint coalitions). This is a major limitation since the space of coalition structures is already exponentially large, and it would be counter-intuitive to search for it in an even bigger space. For example, given 28 agents, the number of coalition structures in $P_{[1,2,3,4,5,6,7]}$ is only $7.8 \times 10^{-9}\%$ of the number of ordered sets in $T_{[1,2,3,4,5,6,7]}$. Note that the difference in size between the two spaces grows exponentially with the number of agents involved.

2. Although NCP does not generate the same ordered set twice, it generates multiple ordered sets containing the same coalitions, but ordered differently. For example, given $A = \{a_1, \cdots, a_7\}$ and $G = [2, 2, 3]$, NCP generates the following ordered sets, $\langle\{a_1, a_2\}, \{a_3, a_4\}, \{a_5, a_6, a_7\}\rangle$ and $\langle\{a_3, a_4\}, \{a_1, a_2\}, \{a_5, a_6, a_7\}\rangle$, which correspond to the same coalition structure. Note that we need to find the best coalition structure and, in order to do so, it is sufficient to examine the value of every coalition structure once. In other words, any operation that results in the same coalition structure being generated more than once is considered redundant.

What would be desirable, then, is to find a way to cycle through the lists $L_{g_1}, \ldots, L_{g_{|G|}}$ such that only valid combinations are generated. In other words, it would be desirable if $C_k$ only cycles through the valid coalitions in $L_{g_k}$, rather than going through every coalition in $L_{g_k}$ and verifying whether it overlaps with $\{C_1, \ldots, C_{k-1}\}$. Moreover, in order to avoid performing any redundant operations, it would be desirable if the cyclation process is guaranteed not to go through the same coalition structure more than once. Algorithm 4 describes a novel cyclation process that meets these requirements.

The basic idea is to use the `searchList` function to cycle through the coalitions in $L_{g_1}$. For each of these coalitions, `searchList` is called recursively[17] to cycle through the coalitions in $L_{g_2}$ that do not overlap with the first coalition (i.e. the one taken from $L_{g_1}$). Similarly, while cycling through $L_{g_2}$, `searchList` is called recursively to cycle through the coalitions in $L_{g_3}$ that do not overlap with the first two coalitions, and so on. This is repeated until `searchList` is called to cycle through the coalitions in $L_{g_{|G|}}$, in which case we have a valid coalition structure (denoted $\overrightarrow{CS}$ in Algorithm 4) that belongs to $P_G$. Then, if $\overrightarrow{CS}$ has a value that is greater than $V(CS')$ then $CS'$ is updated accordingly. The remainder of this section describes how Algorithm 4 avoids generating invalid or redundant coalition structures without making *any* comparison between coalitions. It also describes how the algorithm applies a branch-and-bound technique to speed up the search.

**Avoiding invalid coalition structures:** Given $G = [g_1, \ldots, g_{|G|}]$, we define the following ordered sets of agents: $A_1, A_2, \cdots, A_{|G|}$, where $A_1$ contains $n$ agents, and $A_k : 2 \leq k \leq |G|$ contains $n - \sum_{i=1}^{k-1} g_i$ agents. Moreover, we assume that the agents in $A_k : 1 \leq k \leq |G|$ are ordered

17. `searchList` is not actually implemented in our code as a recursive function (due to the inefficiency of recursive functions in general). However, to make Algorithm 4 easier to understand, the recursive form of the algorithm is presented in the paper.

**Algorithm 4 :** `searchList`$(G, k, \alpha, A_k, CS', \overrightarrow{CS})$ – search a sub-space.

**Require:** $\{max_s\}_{s\in\{1,\cdots,n\}}, \{\mathbf{v}(L_s)\}_{s\in\{1,\cdots,n\}}, UB^*, \beta^*$

1: **if** $k > 1$ and $g_k \neq g_{k-1}$ `{if the size is not being repeated.}` **then**
2:     $\alpha \leftarrow 1$ `{reset` $\alpha$`.}`
3: **end if**
4: **for** $M_k \in LC^{|A_k|}_{g_k}$ such that $\alpha \leq M_{k,1} \leq n + 1 - \sum_{i=1}^{k}(g_k \times G(g_k))$ **do**
5:     $C_k \leftarrow \{A_{k,i} \mid i \in M_k\}$ `{extract` $C_k$ `given` $M_k$ `and` $A_k$`.}`
6:     **if** $k = |G|$ and $V(CS') < V(\overrightarrow{CS})$ **then**
7:         $CS' \leftarrow \overrightarrow{CS}$ `{update the current best.}`
8:     **else if** $V(CS') < \sum_{s\in\{g_1,\cdots,g_k\}} v(C_s) + \sum_{s\in\{g_{k+1},\cdots,g_n\}} max_s$ `{branch only if there is potential of finding a coalition structure better than` $CS'$`.}` **then**
9:         $CS' \leftarrow$ `searchList`$(G, k+1, M_{k,1}, A \setminus C, CS', \overrightarrow{CS})$ `{branch to next coalition.}`
10:     **end if**
11:     **if** $\frac{UB^*}{V(CS')} \leq \beta^*$ or $V(CS') = MAX_G$ `{stop if the required solution has been found or if the current best is equal to the upper bound of this sub-space.}` **then**
12:         **return** $CS'$
13:     **end if**
14: **end for**
15: **return** $CS'$

ascendingly based on their indices in $A$ (e.g. if $A_k$ contains agents $a_5$, $a_7$, and $a_2$, then the order would be $A_k = \langle a_2, a_5, a_7\rangle$). In other words, we assume that: $A_{k,1} < A_{k,2} < \cdots < A_{k,|A_k|}$, where $A_{k,i}$ is the $i^{th}$ agent in $A_k$.[18] Now, given a number of coalitions $C_1, C_2, \cdots, C_{g_{k-1}}$ taken from the lists $L_{g_1}, L_{g_2}, \cdots, L_{g_{k-1}}$ respectively, we show how to cycle through the coalitions in $L_{g_k}$ that do not overlap with any of the aforementioned ones, and that is without storing $L_{g_k}$ in memory. In more detail, this can be done using the following modifications over NCP:

- Instead of using $M_k$ to cycle through the combinations in $LC^n_{g_k}$ (as in NCP), we use it to cycle through the combinations in $LC^{|A_k|}_{g_k}$.
- For any given instance of $M_k$, we extract the corresponding coalition $C_k \in L_{g_k}$ using the following operation: $C_k = \{A_{k,i} \mid i \in M_k\}$ (see step 5 of Algorithm 4). For example, given $M_k = \{1, 3, 5\}$, the corresponding coalition does not contain agents $a_1$, $a_3$, and $a_5$ (as in NCP). Instead, it contains the $1^{st}$, the $3^{rd}$, and the $5^{th}$ element of $A_k$.

These differences ensure that $M_k$ cycles through all the possible coalitions of size $g_k$ that are taken from $A_k$ (instead of those taken from $A$). Based on this, if we set $A_k = A\backslash\{C_1, \cdots, C_{k-1}\}$, then we ensure that every instance of $C_k$ does not overlap with any of the coalitions $C_1, \cdots, C_{k-1}$.

Figure 6 shows an example given $A = A_1 = \{a_1, a_2, a_3, a_4, a_5, a_6, a_7\}$ and $G = [2, 2, 3]$. As can be seen, having $M_1 = \{1, 6\}$ implies that $C_1$ contains the $1^{st}$ and $6^{th}$ agents in $A_1$ (i.e. it implies

18. Recall that we define an order over the agents in $A$ such that, for any two agents $a_i, a_j \in A$, we have $a_i < a_j$ iff $i < j$. For more details, see Section 2.

that $C_1 = \{a_1, a_6\}$). By knowing the agents that belong to $C_1$, we can then assign $A_2$ to those that do not belong to $C_1$, i.e. $A_2 = \{a_2, a_3, a_4, a_5, a_7\}$ (see how the agents in $A_2$ are ordered based on their indices in $A$). As mentioned earlier, $M_2$ would then cycle through all the possible coalitions of size 2 out of $A_2$, and none of these coalitions would overlap with $C_1$. Similarly, having $M_2 = \{3, 5\}$ implies that $C_2$ contains the $3^{rd}$ and $5^{th}$ elements of $A_2$ (i.e. it implies that $C_2 = \{a_4, a_7\}$), and by knowing the agents that belong to $C_2$, we can then assign $A_3$ to those that do not belong to $C_1$ or $C_2$ (i.e. $A_3 = \{a_2, a_3, a_5\}$), and so on.

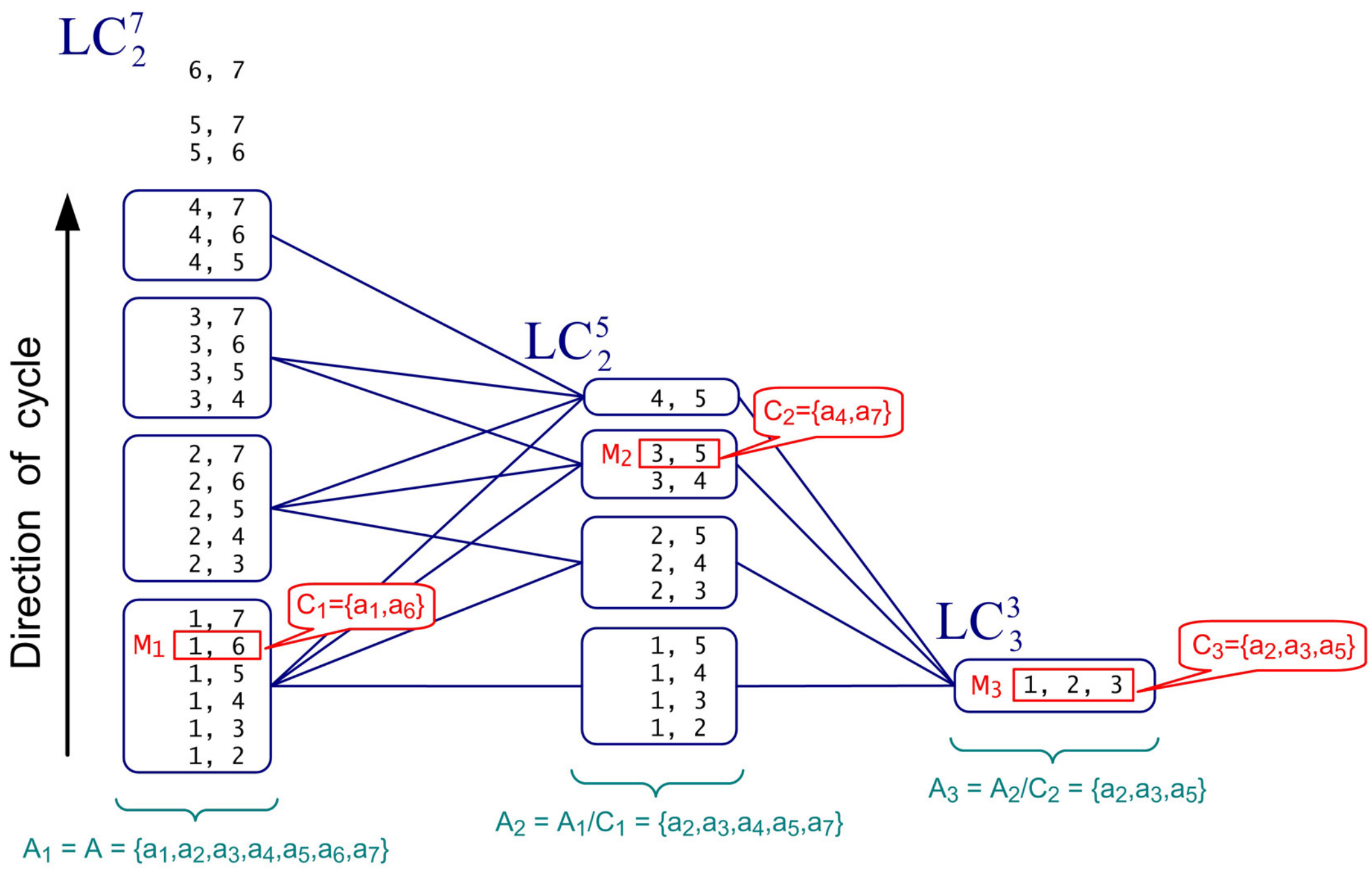

Figure 6: Example of our novel cyclation process, given $A = \{a_1, a_2, a_3, a_4, a_5, a_6, a_7\}$ and $G = [2, 2, 3]$.

The *modified cyclation process* (MCP), which we describe above, generates all the coalition structures in $P_G$ (see Theorem 3), and that is without performing any comparison between the coalitions.

**Theorem 3.** *Given an integer partition $G \in \mathcal{G}$, every coalition structure in $P_G$ is generated by MCP.*

**Proof.** See Appendix D.

Note, however, that MCP suffers from the same limitation of NCP in that it could generate the same coalition structure more than once (e.g. given $G = [2, 2, 3]$, both $\langle\{a_1, a_2\}, \{a_3, a_4\}, \{a_5, a_6, a_7\}\rangle$ and $\langle\{a_3, a_4\}, \{a_1, a_2\}, \{a_5, a_6, a_7\}\rangle$ are generated by MCP). Next, we show how this can be avoided.

**Avoiding redundant coalition structures:** We note that, by using MCP, the same coalition structure can only be generated twice if there are repeated parts in the integer partition $G$ (e.g. $G = [1,2,2,3]$ or $G = [1,4,4,4,6]$). This is because MCP generates ordered sets $\langle C_1, \cdots, C_{|G|} \rangle$ containing disjoint coalitions of which the sizes match the parts in $G$ (i.e. $|C_k| = g_k \in \{1, \cdots, |G|\}$). Based on this, if an ordered set $\overrightarrow{CS}$ is generated by MCP, then, any other ordered set $\overrightarrow{CS}'$ that contains the same coalitions but with a different order (compared to $\overrightarrow{CS}$) will also be generated by MCP as long as the sizes of the coalitions match the parts in $G$. This, of course, can only happen if we have $g_k = g_j : k \neq j$. Based on this, MCP only needs to be modified for the cases where there are repeated parts in $G$.[19] This modification is done as follows:

- While cycling through the combinations in $LC_{g_k}^{|A_k|}$, ensure that the first (i.e. smallest) element in $M_k$ (denoted $M_{k,1}$) satisfies: $\alpha \leq M_{k,1} \leq n+1-\sum_{i=1}^{k}(g_k \times G(g_k))$, where $\alpha = M_{k-1,1}$ if $g_k = g_{k-1}$, and $\alpha = 1$ otherwise (see step 4 of Algorithm 4).

This is illustrated in Figure 6 using the connected boxes. In more detail, $M_1$ only cycles through the combinations in $LC_2^7$ that are contained in boxes (e.g. it does not cycle through combinations $\{5,6\}$, $\{5,7\}$, and $\{6,7\}$). Moreover, $M_2$ only cycles through the combinations in $LC_2^5$ that are contained in boxes connected to the one in which $M_1$ is currently cycling. This modification ensures that $M_{k+1,1} \geq M_{k,1}$ when $g_{k+1} = g_k$. For example, while $M_1$ is cycling through the box in $LC_2^7$ containing the combinations in which the smallest element is 3, we have $M_{1,1} = 3$. In this case, $M_2$ only cycles through the boxes in $LC_2^5$ containing the combinations in which the smallest element is 3 or 4 (see how these boxes are connected in Figure 6), and this ensures that $M_{2,1} \geq M_{1,1}$.

The *final cyclation process* (FCP), which we describe above, generates every coalition structure in $P_G$ exactly once.

**Theorem 4.** *Given an integer partition $G \in \mathcal{G}$, every coalition structure in $P_G$ is generated exactly once by FCP.*

**Proof.** See Appendix E.

Note, however, that given the exponential size of $P_G$, it would be more desirable if we can avoid generating any coalition structure with no potential of having a value greater than the maximum one found so far. Next, we show how this can be done using a branch-and-bound technique.

**Applying Branch-and-Bound:** As mentioned earlier, when cycling through the coalition structures in $P_G$, we only update $C_k$ once we have examined all the possible instances of $\{C_{k+1}, \ldots, C_{|G|}\}$ that do not overlap with $\{C_1, \ldots, C_k\}$. In other words, we only update $C_k$ once we have examined all the possible coalition structures that start with $\{C_1, \ldots, C_k\}$. However, if we knew that none of these coalition structures could have a value greater than the maximum value found so far, then we could update $C_k$ straight away (i.e. without having to go through any of the possible instances of $\{C_{k+1}, \ldots, C_{|G|}\}$). In order to do so, we calculate an upper bound on the values of the coalitions that can be added to $\{C_1, \ldots, C_k\}$. Specifically, having computed $max_s$ for every possible coalition

19. Note that most of the coalition structures usually contain repeated coalition sizes (e.g. 99.6% of them given 20 agents).

of size $s \in \{1, 2, \ldots, n\}$, we can then calculate such an upper bound, denoted $MAX_{[g_{k+1},\cdots,g_{|G|}]}$, as follows:

$$MAX_{[g_{k+1},\cdots,g_{|G|}]} = \sum_{i=k+1}^{|G|} max_{g_i}$$

Now, if we define $V(\{C_1, \cdots, C_k\})$ as the sum of the values of coalitions $C_1, \cdots, C_k$ (that is, $V(C_1, \ldots, C_k) = \sum_{i=1}^{k} v(C_i)$), then $V(\{C_1, ..., C_k\}) + MAX_{[g_{k+1},\cdots,g_{|G|}]}$ represents an upper bound on the value of the coalition structure that could be obtained with a coalition structure starting with $\{C_1, \cdots, C_k\}$ and ending with coalition sizes $[g_{k+1}, \ldots, g_{|G|}]$.

Hence, having $V(CS') \geq V(\{C_1, ..., C_k\}) + MAX_{[g_{k+1},\cdots,g_{|G|}]}$ implies that none of the coalition structures that start with $\{C_1, ..., C_k\}$ and end with coalitions of sizes: $g_{k+1}, ..., g_{|G|}$ has a value greater than $V(CS')$ (this is checked in step 8 of Algorithm 4). On the other hand, having $V(CS') < V(\{C_1, \ldots, C_k\}) + MAX_{[g_{k+1},\cdots,g_{|G|}]}$ implies that there could be a coalition structure that starts with $\{C_1, \cdots, C_k\}$ and is better than the current best coalition. However, this still does not necessarily imply that *all* of these coalition structures need to be examined. This is because, when the algorithm moves to the next list, it may find that there are certain coalition structures that are not better than the current best. Formally, for every coalition $C_j : k < j < |G|$, we can still have: $V(CS') > V(\{C_1, ..., C_j\}) + MAX_{[g_{j+1},\cdots,g_{|G|}]}$. Figure 7 illustrates how this branch-and-bound technique is applied while searching a sub-space.

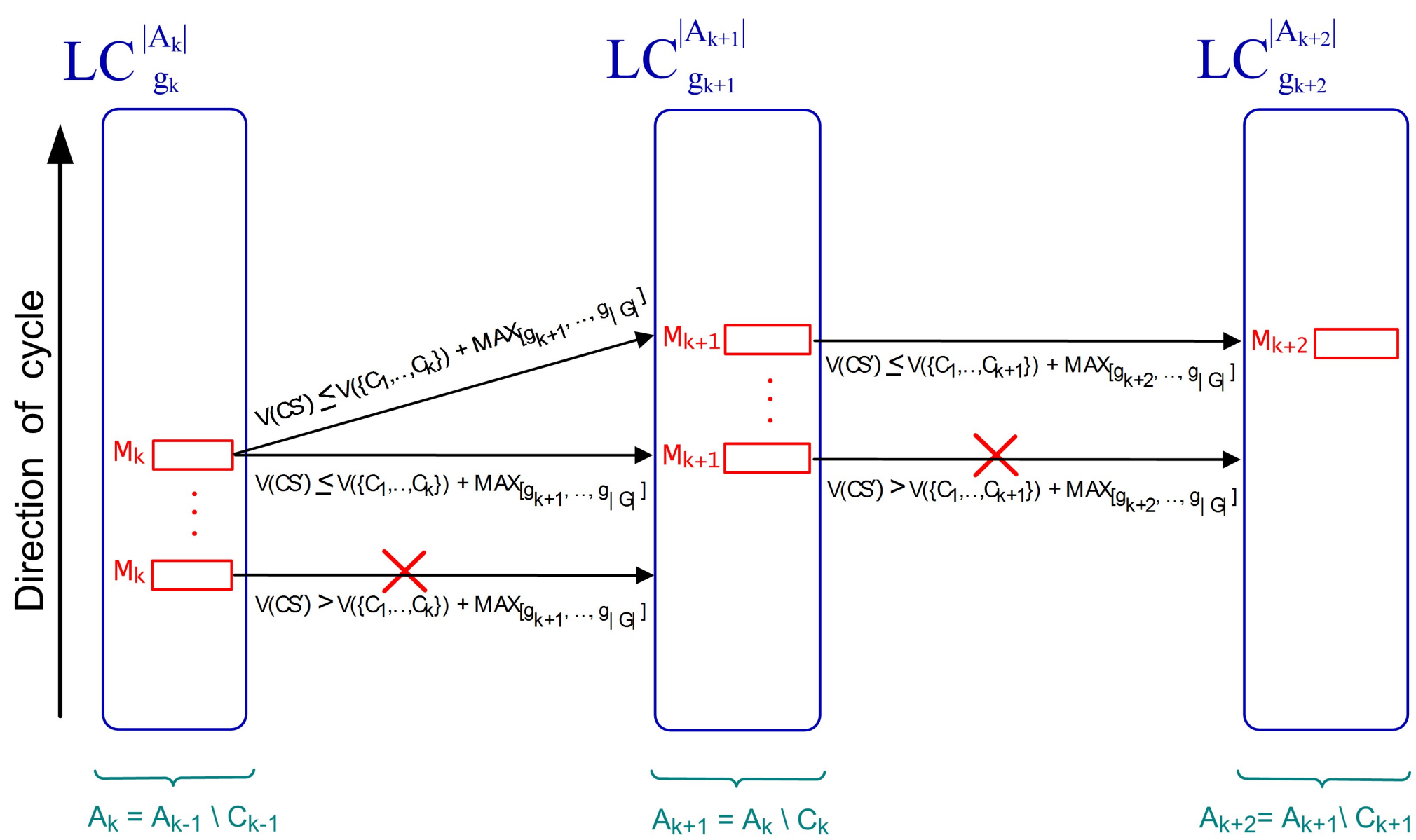

Figure 7: Applying branch-and-bound while searching through the coalition structures in a sub-space.

## 5. Performance Evaluation

In this section, we empirically evaluate the IP algorithm, and benchmark it against the state of the art in the literature. Since IP's ability to prune the space depends on the closeness of the upper and lower bounds to the actual optimal value, and since this closeness is determined by the spread of the distribution of the coalition values, it is crucial that IP is tested against different value distributions. Moreover, we aim to evaluate the ability of our algorithm to generate solutions anytime and to zoom in on very high quality solutions rapidly.

In what follows, we first discuss the validity and properties of the different value distributions that we use to test the algorithm (Section 5.1). Then, we benchmark our algorithm against the fastest available algorithm in the literature (i.e. IDP) using the aforementioned distributions (Section 5.2). Finally, we empirically evaluate the efficiency and effectiveness of our algorithm in generating solutions anytime (Section 5.3).

### 5.1 Benchmarking

The common practice in benchmarking search heuristics is to choose some standard instances of the problem and compare the various algorithms that exist without giving them *a priori* knowledge of the type of input they are presented with. The standard instances for the coalition structure generation problems have been defined and used by Larson and Sandholm (2000) namely:[20]

1. **Normal**: $v(C) \sim |C| \times N(\mu, \sigma^2)$ where $\mu = 1$ and $\sigma = 0.1$.
2. **Uniform**: $v(C) \sim |C| \times U(a, b)$ where $a = 0$ and $b = 1$.

While we use the above distributions to benchmark our algorithm, we also question the validity of these distributions. This is because, in our previous work (Rahwan et al., 2007b), we noted that the normal and uniform distributions tend to generate solutions with small numbers of coalitions. However, we now show that, if the coalition values are picked from the Normal or Uniform distributions (scaled by the size of the coalition), then the resulting distribution of the coalition structure values is biased (see Theorem 5). Given this, experiments defined according to the Normal and Uniform distributions could favour some algorithms over others.

**Theorem 5.** *If the coalition values were taken from a normal distribution as follows:* $\forall C \subseteq A, v(C) \sim |C| \times N(\mu, \sigma^2)$*, or if they were taken from a uniform distribution as follows:* $\forall C \subseteq A, v(C) \sim |C| \times U(a, b)$*, then, given any coalition structure* $CS' : |CS'| > 1$*, there exists another coalition structure* $CS'' : |CS''| < |CS'|$ *such that:*

$$P\left(V(CS'') = V(CS^*)\right) > P\left(V(CS') = V(CS^*)\right)$$

*That is, the probability of* $CS''$ *being an optimal coalition structure is greater than that of* $CS'$*.*

**Proof.** See Appendix F.

20. Their sub and super-additive distributions are also studied in the literature, but in such cases it is usually known *a priori* that the distribution of coalition values is actually of these types (in which case it is known *a priori* what the optimal coalition structure is). Moreover, previous results on these distributions have not produced very interesting insights (Rahwan et al., 2007b) and so we do not experiment with these.

To remedy this, we propose a new input distribution that is tailored specifically to the CSG problem. This distribution, which we define as NDCS (Normally Distributed Coalition Structures), is constructed by generating coalition values in the following way:

**NDCS**: $v(C) \sim N(\mu, \sigma^2)$, where $\mu = |C|$ and $\sigma = \sqrt{|C|}$.

In this case, it turns out that the value of every possible coalition structure is independently drawn from the *same* normal distribution which leads us to the following theorem:

**Theorem 6.** *Iff we have: $\forall C \subseteq A, v(C) \sim N(\mu, \sigma^2)$, where $\mu = |C|$ and $\sigma = \sqrt{|C|}$, then the following holds:*

$$\forall CS \in \mathcal{P}, V(CS) \sim N(|A|, |A|)$$

**Proof.** See Appendix G.

Since the NDCS distribution ensures that every coalition structure value is drawn from the same distribution, it ensures that the search space is not biased. Thus, the efficiency of search algorithms in finding the optimal coalition structure is more strongly tested than in the other cases.

Using the above input distributions, we benchmark our algorithm against the other state-of-the-art algorithm, namely IDP (see Section 2). Note that we do not experiment with the other anytime algorithms since they need to search the whole space to find the optimal value and this is generally not feasible within reasonable time, even for small numbers of agents. Also, it was shown by Rahwan et al. (2007b) that industrial strength software such as CPLEX cannot handle inputs of more than 18 agents since it runs out of memory and therefore we do not run experiments with it here. On all our graphs we plot the 95% confidence interval at every point (given 800 runs for 15 to 20 agents and 100 runs for 21 to 25 agents).[21]

### 5.2 Experiment 1: Optimality

In this experiment, we compare the algorithms' performances given different numbers of agents (from 15 to 27). The time to find the optimal coalition structure is measured in terms of clock time (in milliseconds) on an Intel 2.6GHz Quad Core PC with 3Gigabytes of RAM. The algorithms are coded using JAVA 1.6. The running times are plotted on a log scale in Figure 8.[22] We note as IP-X the application of IP to distribution X, where X can be NDCS, Normal, or Uniform (as described above). As can be seen, IP finds the optimal coalition structure significantly faster than IDP for all distributions. In the best case (Uniform for 27 agents) IP is 570 times better than IDP (i.e. it takes 0.175% of the time taken by IDP) and in the worst case (NDCS for 16 agents) it is 1.7 times faster than IDP. It can also be seen that the performance of IP is the slowest given the NDCS distribution (compared to IP-Normal and IP-Uniform). To determine the cause for this, we first discuss the two main problems that can affect the performance of IP:

21. By plotting the 95% confidence interval, we aim to check statistical significance of the difference between the means taken at each point across different series. Thus, if two points from two different series have overlapping confidence intervals, it is equivalent to saying that the null hypothesis is validated (i.e. the means are not significantly different) for a t-test with $\alpha = 0.05$. If the confidence intervals do not overlap, then the means are significantly different.

22. The running time for IDP is deterministic since it runs in $O(3^n)$. Hence, we recorded its running time for up to 25 agents and extrapolated the results to 27 agents.

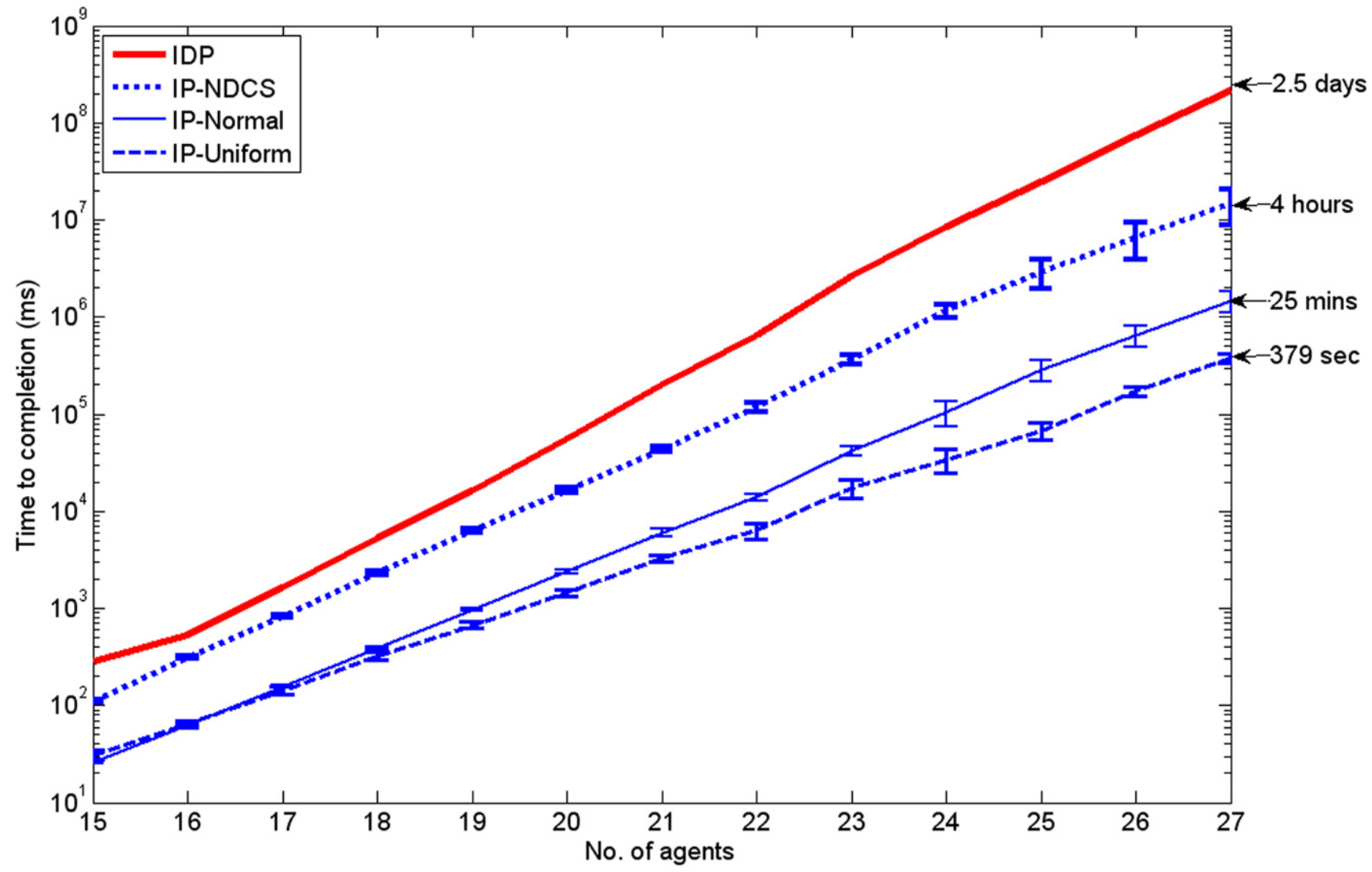

Figure 8: Time to find the optimal solution for IDP, IP applied to NDCS, Normal, and Uniform distributions.

1. Pruning sub-spaces: the higher the upper bounds of sub-spaces and the lower the value of the optimal coalition structure, the harder it is to prune sub-spaces. This can be deduced from the pruning function we use in Algorithm 2. Moreover, the bigger the sub-spaces with higher upper bounds, the longer the algorithm will take to find the optimal solution. This is because the algorithm always has to search the sub-space with the highest upper bound to check that the solution it has found is optimal.

2. Branch-and-bound: the higher the upper bounds of sub-spaces and the lower the optimal coalition structure value, the harder it is to prune with branch-and-bound. This can be deduced from the pruning applied in step 8 of Algorithm 4. This is because, when applying branch-and-bound within a sub-space $P_{\{g_1,g_2,\ldots,g_n\}}$, the current best solution $CS'$ is compared against the sum of coalition values and the maximum value of coalitions of the remaining coalition sizes as follows:

   $$\textbf{if } V(CS') > \sum_{C \in \{L_{g_1},\cdots,L_{g_k}\}} v(C) + \sum_{g \in \{g_{k+1},\cdots,g_n\}} max_g \textbf{ then } \text{move to next coalition structure}$$

   Now, if the best solution is very low compared to the upper bound, that is:

   $$V(CS') << UB_G = \sum_{g \in \{g_1,\cdots,g_n\}} max_g$$

   then, branch-and-bound has to be applied deeper (i.e. increasing variable $k$ in the condition above) in the sub-space in order to make sure that the coalition structure being evaluated is not optimal. Hence, in the worst case it would have to search the whole sub-space (i.e. apply step 9 in Algorithm 4 on increasing values of $k$ up to $n$).

In order to see how these different issues affect the performance of our algorithm with respect to different distributions, we recorded the value of the optimal coalition structure and the upper bounds of all the sub-spaces (given 21 agents) and averaged them over 20 runs.[23] We also exactly recorded the size of each sub-space (i.e. in the number of coalition structures per sub-space). The results are plotted in Figure 9. We note the following for each distribution:

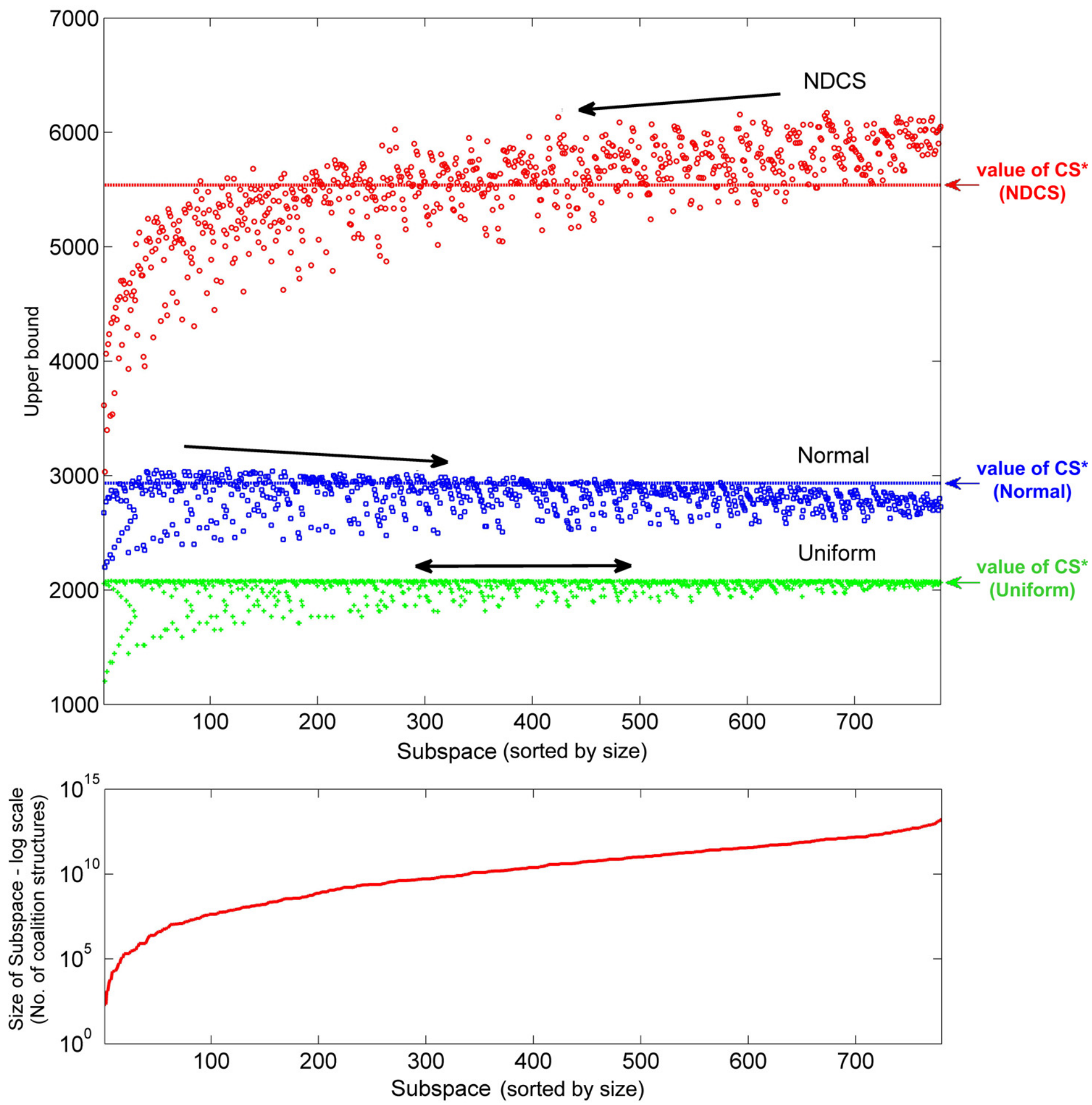

Figure 9: Top: upper bounds and optimal coalition structure value, bottom: size of sub-spaces. Note that the values in the bottom graph are plotted on a log scale. Points with the same abscissa on the two graphs correspond to the same sub-space. The arrows show the direction of the search for each distribution.

- NDCS: The biggest sub-spaces are the ones with the highest upper bounds. Hence, it is much harder to prune large portions of the search space. Moreover, the average optimal

23. The values of the upper bounds and the average optimal coalition structure were rounded and scaled to ease the explanation and to have a clearer plot.

coalition structure value is relatively low compared to the upper bounds of the bigger sub-spaces. Hence, applying branch-and-bound in this distribution is very hard.

- Normal: The smaller sub-spaces are the ones with the highest upper bounds. Hence, pruning large portions of the space can easily be done by searching smaller sub-spaces in which good solutions are. Moreover, the value of the optimal coalition structure tends to be higher than the upper bounds of most large sub-spaces, and relatively close to the highest upper bounds. Hence, it is easier for branch-and-bound to prune large portions of the sub-spaces.

- Uniform: The upper bounds of *most* sub-spaces are relatively high compared to those of other distributions (i.e. they are close to the highest upper bound). In fact, the upper bounds are actually *nearly equal* to the average optimal solution and this allows the algorithm to prune most of the sub-spaces as soon as it has found an optimal solution, and this happens almost immediately after scanning the input.

Finally, note that Figure 9 shows the portion of the space that will be avoided given the selection strategy described earlier in Section 4.2. In more detail, recall that this strategy is guaranteed to avoid searching the sub-spaces that have an upper bound lower that $V(CS^*)$. As can be seen from the figure, many of the sub-spaces (in the case of NDCS and Uniform distributions) have an upper bound lower than $V(CS^*)$, although most of these sub-spaces are relatively small. Moreover, in the case of the Normal distribution, almost all the sub-spaces have an upper bound lower than $V(CS^*)$, most of which are among the largest ones!

Having studied the performance of IP in terms of completion time, we next focus on studying its ability to generate solutions anytime.

### 5.3 Experiment 2: Anytime Quality

In this experiment, we further evaluate the anytime property of our algorithm, and that is by recording the value of the solutions that were generated before returning the guaranteed optimal one. In particular, we recorded two indicative measures of the quality of the solutions. First, we computed the ratio between the value of the current best solution and the optimal solution (obtained at the end of the run). This ratio is noted as $r_{opt} = \frac{V(CS')}{V(CS^*)}$. This measure shows how effective the algorithm is at zooming on good solutions. Second, we recorded the ratio $r_{bound}$ between the value of the current best solution and the upper bound on the optimal value (i.e. $r_{bound} = \frac{V(CS')}{UB^*}$). This measure is the theoretical guarantee that the algorithm places on the quality of the solution (see Section 4.1). Ideally, the algorithm should be able to minimise the difference between $r_{opt}$ and $r_{bound}$ in minimal time.

The results are plotted in Figure 10 for the distributions: NDCS, Normal, and Uniform.[24] We discuss the results for each of the distributions in turn.

- NDCS: As can be seen, the algorithm very high quality guarantees (i.e. $r_{bound} > 90\%$) in less than half of the time required to find the optimal solution. It also produces a very high quality solutions (i.e. $r_{opt} > 90\%$) within less than $10\%$ of the time required to terminate.

24. The points plotted are averages computed over 500 runs for 19 agents and 22 agents, and 100 runs for 25 agents. The error bars depict the $95\%$ confidence interval for each of the intervals over which results are recorded.

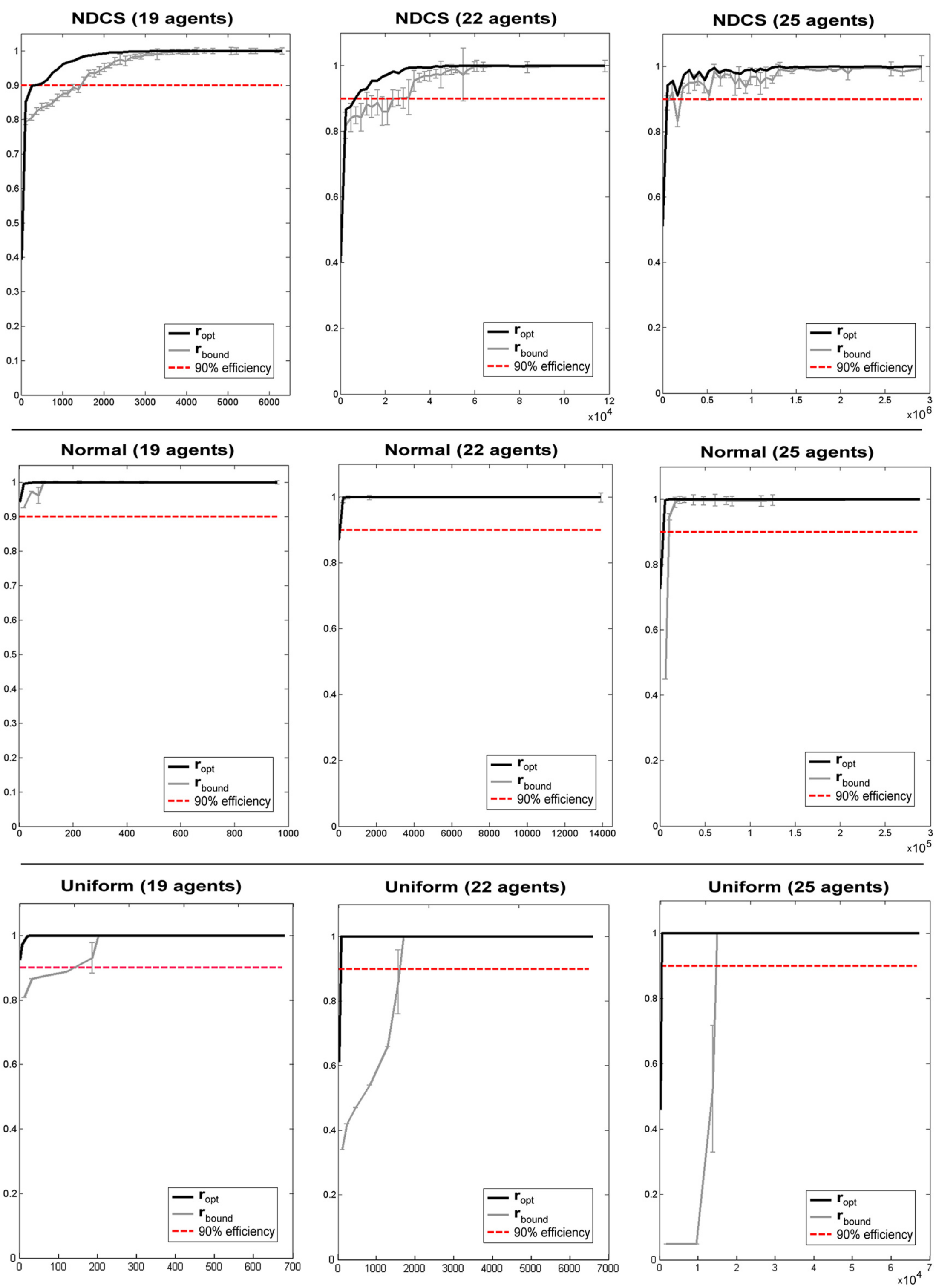

Figure 10: Quality ($r_{opt}$) and bound ($r_{bound}$) for the generated solution. In all cases, the x-axis represents the time (in milliseconds) and the y-axis represents the ratio of the solution to the optimal.

- Normal: In this case, our algorithm is able to come up with guaranteed high quality solutions much faster than for the NDCS distribution. Moreover, in this case, very high quality solutions (i.e. $r_{opt} > 90\%$) *can be guaranteed* (i.e. $r_{bound} > 90\%$) in less than $10\%$ of the time to find the optimal value. This results from the fact that the upper bounds are not as far from the optimal value as in the NDCS case.

- Uniform: As expected from earlier results presented in Section 5.2, the algorithm generates very high quality solutions (i.e. $r_{opt} \approx 100\%$) faster than for the other distributions (shortly after scanning the input). Moreover, the solutions can be guaranteed to be near-optimal (i.e. $r_{bound} > 99\%$) within $15\%$ of the time to find the optimal.

Next, we compare the worst-case guarantees that are provided by IP with those provided by Sandholm et al.'s (1999) and Dang and Jennings's (2004) algorithms (see Figure 11). As can be seen, our algorithm significantly outperforms both Dang and Jennings's and Sandholm et al.'s for all distributions. In particular, after scanning the input, IP is able to guarantee that its solution is nearly 40% (in the worst case) of the optimal compared to below 10% for the other algorithms. Moreover, our guarantee usually reaches 100% after searching minute portions of the search space (on average around 0.0000019% for the hardest distribution), while the guarantees provided by other algorithms do not go beyond 50% until the whole space has been searched. Also note that we generate very high quality solutions (i.e. $> 90\%$) by searching even smaller portions of the of the search space (on average around 0.0000002% for the hardest distribution). Thus, in actual computational time, for 25 agents for example, we are able to return a solution that is guaranteed to be higher than $90\%$ of the optimal in around 250 seconds in the worst case and 300 milliseconds in the best case.

## 6. Conclusions and Future Work

Coalition formation, the process by which a group of software agents come together and agree to coordinate and cooperate in the performance of a set of tasks, is an important form of interaction in multi-agent systems. Such coalitions can improve the performance of the individual agents and/or the system as a whole, especially when tasks cannot be performed by a single agent, or when a group of agents performs the tasks more efficiently. One of the most challenging problems that arise in the coalition formation process is that of coalition structure generation, which involves partitioning the set of agents into exhaustive and disjoint coalitions such that the social welfare is maximized. In this paper, we have developed and evaluated an anytime integer-partition based algorithm (called IP) that finds optimal solutions much faster than any previous algorithm designed for this purpose. The strength of our approach is founded upon two main components:

- We use a novel representation of the search space which partitions it into smaller, disjoint sub-spaces that can be explored independently to find optimal solutions. This representation, which is based on the integer partitions of the number of agents involved, allows the agents to balance the trade-offs between their preferences for certain coalition sizes against the computation required to find the solution. Moreover, such trade-offs can be made in an informed manner since we can compute bounds on sub-spaces of the search space. These bounds allow us to prune the search space and guarantee the quality of the solution found during the search. They may also, depending on the distribution of the input values, allow us to obtain the optimal solution almost immediately after scanning the input.

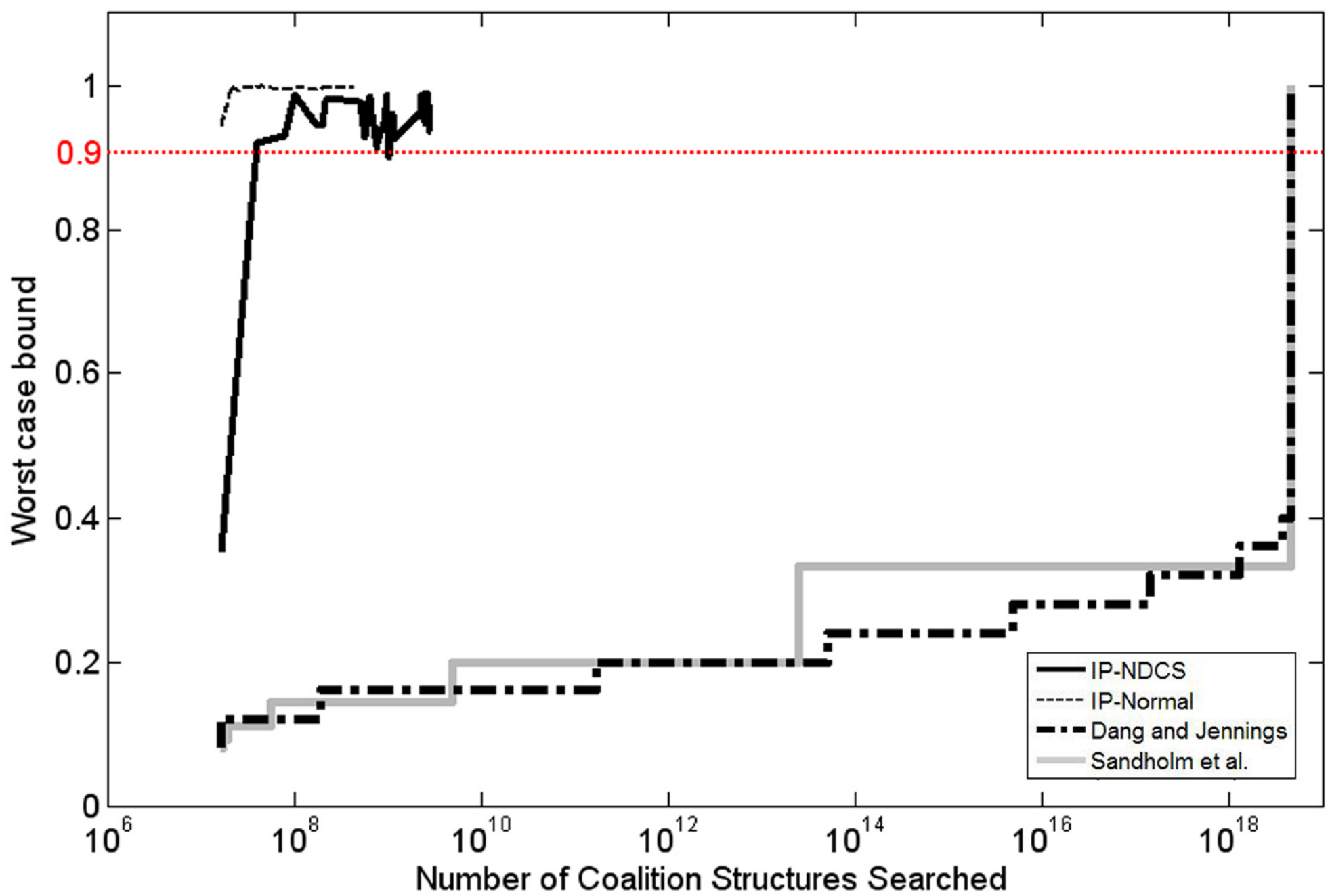

Figure 11: Worst case bounds generated by IP using the Normal and NDCS distributions compared to Sandholm et al.'s (1999) and Dang and Jennings's (2004) algorithms for 25 agents. The results for the Uniform distribution are trivial since IP on average finds the optimal almost immediately after scanning the input. Note that error bars have been omitted from the IP results for reasons of clarity.

- We devise a technique that allows us to cycle through the coalition structures within a given sub-space. Unlike a naïve cyclation technique that generates combinations of coalitions, and verifies whether each of these combinations is a valid coalition structure, our cyclation technique only generates valid ones (thus, avoiding the search through the space of possible combinations of coalitions, which is exponentially larger than the space of coalition structures). In addition, the cyclation technique does not perform any redundant operations since it avoids generating the same coalition structure more than once. Finally, by applying a branch-and-bound technique, we are able to identify the coalition structures that cannot improve on the quality of the solution found so far, and thus, avoid generating them.

Altogether, these components allow us to make significant performance gains over other existing approaches. In more detail, the experiments show that IP avoids searching most of the search space, and therefore, requires significantly less time, compared to the other algorithms, in order to return an optimal solution. For example, IP outperforms IDP by orders of magnitude ($0.175\%$ of the time taken by IDP for 27 agents in the best case). Moreover, if IP is interrupted before an optimal value is found, it can still return solutions that are very close to the optimal (usually above $95\%$ of the optimal), with very high worst-case guarantees on them (usually above $90\%$). These solutions are always better (above $40\%$ of the optimal right after scanning the input) than those returned by Sandholm et al.'s (1999) and Dang and Jennings's (2004) algorithms (i.e. less than $10\%$ of the optimal). These algorithms also have to search a large portion of the search space before being able to get better guarantees while our algorithm is able to prune and find near-optimal solutions relatively quickly (above $90\%$ of the optimal within 10% of the time to find the optimal solution for 25 agents).

A number of important extensions to IP could be envisaged. For example, we have recently combined the IDP algorithm with IP (IP-IDP) (Rahwan & Jennings, 2008a) and will explore other approaches including linear programming techniques to improve the bounds used in IP. However, these extensions have to deal with an exponential input (i.e. $2^n$ memory locations at least for $n$ agents) as we do in IP. Therefore, it is important to develop techniques that will extend our approach in order to minimise cycling through all coalition values as the number of agents increases. This will require adapting our cyclation technique and the bound computation. Hence, in future work, we will need to devise representations for sub-spaces that allow us to cycle more intelligently over larger inputs and develop new techniques to compute bounds to be used by our branch-and-bound algorithm. In trying to adapt our approach to other problems, we also aim to determine the degree to which IP can be used to solve other common incomplete set partitioning problems which occur in combinatorial auctions (Rothkopf et al., 1995) or crew scheduling (Hoffman & Padberg, 1993). Finally, we aim to see whether the patterns that we exploit in our algorithm also arise in other combinatorial optimisation problems that have been studied in the area of combinatorics (e.g., Kreher & Stinson, 1998; Papadimitriou & Steiglitz, 1998).

## 7. Acknowledgments

The research in this paper was undertaken as part of the ALADDIN (Autonomous Learning Agents for Decentralised Data and Information Systems) project and is jointly funded by a BAE Systems and EPSRC (Engineering and Physical Research Council) strategic partnership (EP/C548051/1). Andrea Giovannucci was funded by the Juan de la Cierva programme (JCI-2008-03006) and the EU funded Synthetic Forager project (ICT-217148-SF). We also wish to thank Professor Tuomas

Sandholm for his comments, as well as the anonymous reviewers for their valuable comments on previous versions of the paper. We are also very grateful to Dr. Viet Dung Dang for his contributions to earlier versions of the paper. Finally, we wish to thank to Dr. W. T. Luke Teacy for his help with some of the proofs and the anonymous reviewers for their very constructive comments.

## Appendix A. Summary of Notation

| | |
|---|---|
| $A$ | The set of agents. |
| $a_i$ | An agent in $A$. |
| $n$ | The number of agents in $A$. |
| $C$ | A coalition. |
| $\|C\|$ | The cardinality of $C$. |
| $v(C)$ | The value of $C$. |
| $CS$ | A coalition structure. |
| $V(CS)$ | The value of $CS$. |
| $CS^*$ | An optimal coalition structure. |
| $UB^*$ | The upper bound on $V(CS^*)$. |
| $LB^*$ | The lower bound on $V(CS^*)$. |
| $CS'$ | The best coalition structure found so far. |
| $\beta$ | The bound on the quality of the best solution found so far. |
| $\beta^*$ | The bound within which any solution is acceptable. |
| $LC_s^i$ | The list of possible combinations of size $s$ taken from the set $\{1, 2, \ldots, i\}$. |
| $L_s$ | The list of coalitions of size $s$ ordered lexicographically. |
| $\mathbf{v}(L_s)$ | A list containing the values of all the coalitions in $L_s$. |
| $max_s$ | The maximum value of the coalitions in $L_s$. |
| $min_s$ | The minimum value of the coalitions in $L_s$. |
| $avg_s$ | The average value of the coalitions in $L_s$. |
| $\mathcal{P}$ | The set of possible coalition structures. |
| $\mathcal{P}_i$ | The $i^{th}$ level in our representation of the space of possible coalition structures. |
| $LV_i$ | The $i^{th}$ level of the coalition structure graph. |
| $G$ | An integer partition of $n$. |
| $G(s)$ | The multiplicity of $s$ in $G$. |
| $E(G)$ | The underlying set of $G$. |
| $\mathcal{G}$ | The set of possible integer partitions of $n$. |
| $\mathcal{G}^2$ | The set of possible integer partitions of $n$ that contain two parts each. |
| $T_G$ | The Cartesian product of the lists $L_s : s \in G$. |
| $P_G$ | The sub-space (in our space representation) that corresponds to $G$ (i.e. the pre-image of $G$ under $F$). |
| $MAX_G$ | The maximum value of the elements in $T_G$. |
| $MIN_G$ | The minimum value of the elements in $T_G$. |
| $AVG_G$ | The average value of the elements in $T_G$. |
| $F$ | A function that maps a coalition structure $CS$ to an integer partition $G$ such that: $\forall C \in CS, \exists g \in G : \|C\| = g$. |
| $\overrightarrow{CS}$ | A variable used to cycle through the coalition structures in $P_G$. |
| $M_k$ | A variable used to cycle through a list of combinations of size $g_k$. |
| $A_k$ | An ordered set containing the agents that are not members of $C_1, \ldots, C_{k-1}$. |
| $C_s^n$ | The binomial coefficient (i.e. the number of possible combinations of size $s$ taken from $n$ elements). |
| $P(x)$ | The probability of $x$. |
| IDP | The improved dynamic programming algorithm. |
| $N(\mu, \sigma^2)$ | Normal distribution with mean $\mu$ and variance $\sigma^2$. |
| $U(a, b)$ | Continuous Uniform distribution on the interval $[a, b]$. |
| $r_{opt}$ | The ratio between the value of the current best solution and the value of the optimal solution. |
| $r_{bound}$ | The ratio between the value of the current best solution and the upper bound on the value of the optimal solution. |

## Appendix B. Proof of Theorem 1.

Let $\bar{G} = [g_1, g_2, \ldots, g_{|G|}]$ contain the elements of $G$ with a natural ordering on them, and let $P_{\bar{G}}$ return all *ordered coalition structures* $\langle C_1, C_2, \ldots, C_{|G|} \rangle$: $C_i \in L_{g_i}$, where the order of the coalitions within the coalition structure is taken into consideration. For example, given $n = 4$ and $G = [1, 1, 2]$, we have two ordered coalition structures: $\langle \{a_1\}, \{a_2\}, \{a_3, a_4\} \rangle$ and $\langle \{a_2\}, \{a_1\}, \{a_3, a_4\} \rangle$ in $P_{\bar{G}}$ that correspond to one coalition structure: $\{\{a_1\}, \{a_2\}, \{a_3, a_4\}\}$ in $P_G$. Now, since the number of repetitions of each coalition structure in $P_{\bar{G}}$ is the same[25] (e.g., in the above example with $\bar{G} = [1, 1, 2]$, all coalition structures in $P_G$ will appear twice in $P_{\bar{G}}$), then we have:

$$AVG_G = AVG_{\bar{G}} \tag{3}$$

where $AVG_{\bar{G}}$ is the average value of the coalition structures in $P_{\bar{G}}$. Now, if we define $N_n(g_1, g_2, \ldots, g_{|G|})$ as the number of ordered coalition structures in $P_{\bar{G}}$, then we have:

$$\begin{aligned} AVG_{\bar{G}} &= \frac{1}{N_n(g_1, g_2, \ldots, g_{|G|})} \sum_{CS \in P_{\bar{G}}} V(CS) \\ &= \frac{1}{N_n(g_1, g_2, \ldots, g_{|G|})} \sum_{CS \in P_{\bar{G}}} \sum_{C \in CS} v(C) \end{aligned}$$

Moreover, for every coalition $C \in L_{g_i}$, there are: $N_{n-g_i}(g_1, g_2, \ldots, g_{i-1}, g_{i+1}, \ldots, g_{|G|})$ ordered coalition structures where $C$ happens to be the $i^{th}$ coalition. Based on this, we have:

$$N_n(g_1, g_2, \ldots, g_{|G|}) = |L_{g_i}| \times N_{n-g_i}(g_1, \ldots, g_{i-1}, g_{i+1}, \ldots, g_{|G|}) \tag{4}$$

Similarly, the number of times that $v(C)$ occurs in the $i^{th}$ position of the sum of all coalition values in $P_{\bar{G}}$ is $N_{n-g_i}(g_1, \ldots, g_{i-1}, g_{i+1}, \ldots, g_{|G|})$. Given this, we next compute $AVG_{\bar{G}}$ as follows:

$$\begin{aligned} AVG_{\bar{G}} &= \frac{1}{N_n(g_1, g_2, \ldots, g_{|G|})} \sum_{i=1}^{|G|} \sum_{C \in L_{g_i}} N_{n-g_i}(g_1, \ldots, g_{i-1}, g_{i+1}, \ldots, g_{|G|}) \times v(C) \\ &= \sum_{i=1}^{|G|} \sum_{C \in L_{g_i}} \frac{N_{n-g_i}(g_1, \ldots, g_{i-1}, g_{i+1}, \ldots, g_{|G|})}{N_n(g_1, g_2, \ldots, g_{|G|})} \times v(C) \\ &= \sum_{i=1}^{|G|} \sum_{C \in L_{g_i}} \frac{1}{|L_{g_i}|} \times v(C) \text{ (following equation (4))} \\ &= \sum_{i=1}^{|G|} \left( \frac{1}{|L_{g_i}|} \sum_{C \in L_{g_i}} \times v(C) \right) \\ &= \sum_{i=1}^{|G|} avg_{g_i} \end{aligned}$$

25. Specifically, a coalition structure is repeated $x!$ times if it contains $x$ coalitions of the same size.

Based on this, as well as (3), we find that:

$$AVG_G = \sum_{i=1}^{|G|} avg_{g_i}$$

$\square$

## Appendix C. Proof of Theorem 2.

Generally speaking, given a set $B \subseteq A$, the number of possible combinations of size $s$ out of the set $\bar{B} = A \backslash B$ is equal to $C_s^{|A|-|B|}$. Based on this, for every coalition $C$ of size $g_1$, there are $C_{g_2}^{n-g_1}$ coalitions of size $g_2$ that do not overlap with it. Similarly, for every $i$ disjoint coalitions $(C_1, C_2, \ldots, C_i)$ of sizes $g_1, g_2, \ldots, g_i$ respectively, there are $C_{g_{i+1}}^{n-(g_1+g_2+\ldots+g_i)}$ coalitions of size $g_{i+1}$ that do not overlap with the union $C_1 \cup C_2 \cup \ldots \cup C_i$.

Based on this, if $T_G$ is the cartesian product of the lists $L_{g_i} : g_i \in G$, and $\hat{T}_G$ is a subset of $T_G$ that contains only the elements (i.e. the combinations of coalitions) in which no coalitions overlap, then the number of elements in $\hat{T}_G$ can be computed as follows:

$$\left|\hat{T}_G\right| = C_{g_1}^n \times C_{g_2}^{n-g_1} \times \ldots \times C_{g_{|G|}}^{n-(g_1+\ldots+g_{|G|-1})} \tag{5}$$

Moreover, note that any combination of coalitions $\{C_1, C_2, \ldots, C_{|G|}\}$, such that $\forall i \in \{1, 2, \ldots, |G|\}$ : $|C_i| = g_i$, appears exactly once in $P_G$ (since it is considered a unique coalition structure) but could appear more than once in $\hat{T}_G$ (since the ordering of the coalitions matters in the elements of $\hat{T}_G$). In particular, if $g_i$ appears $x$ times in $G$, then every coalition structure in $P_G$ corresponds to $x!$ elements in $\hat{T}_G$, where the coalitions of size $g_i$ are ordered differently in each of these elements. For example, given $G = [1, 2, 2, 2, 5]$, size 2 appears 3 times in $G$ and this means that every coalition structure $\{C_1, C_2, C_3, C_4, C_5\} \in P_G$ corresponds to 3! elements in $\hat{T}_G$ (since 3! is the number of possible permutations of $C_2, C_3, C_4$). This can be generalized as follows:

$$\forall G = [g_1, g_2, \ldots, g_{|G|}] \in \mathcal{G}, |P_G| = \frac{\left|\hat{T}_G\right|}{G(g_1)! \times G(g_2)! \times \ldots \times G(g_{|G|})!} \tag{6}$$

where $G(g_i)$ denotes the multiplicity of $g_i$ in $G$. Then, from (5) and (6), we find that:

$$|P_G| = \frac{C_{g_1}^n \times C_{g_2}^{n-g_1} \times \ldots \times C_{g_{|G|}}^{n-(g_1+\ldots+g_{|G|-1})}}{\prod_{s \in E(G)} G(s)!}$$

where $E(G)$ is the underlying set of $G$.

$\square$

## Appendix D. Proof of Theorem 3.

Given an integer partition $G = [g_1, \ldots, g_{|G|}] \in \mathcal{G}$, we need to prove that all the coalition structures in $P_G$ are generated by MCP. Without loss of generality, we will assume that the parts in $G$ are in increasing order. That is:

$$g_1 \leq g_2 \leq \ldots \leq g_{|G|} \tag{7}$$

Now, the way MCP works is by generating ordered sets of coalitions such that, for every ordered set, the first coalition belongs to $L_{g_1}$ and the second belongs to $L_{g_2}$ and so on. Moreover, the way these ordered sets are generated ensures that the coalitions in each of the ordered sets do not overlap. In other words, MCP generates a subset of $T_G$, denoted $\hat{T}_G$, which is defined as follows:[26]

$$\hat{T}_G = \left\{ \left\langle C_1, ..., C_{|G|} \right\rangle \mid \forall i \in \{1, ..., |G|\}, C_i \subseteq A \text{ and } |C_i| = g_i \text{ and } \forall j \in \{1, ..., |G|\} : j \neq i, C_i \cap C_j = \emptyset \right\}$$

Then, given a coalition structure $CS \in P_G$, let $\hat{T}_G^{CS}$ be a subset of $\hat{T}_G$ containing all the ordered sets that correspond to $CS$. That is:

$$\hat{T}_G^{CS} = \left\{ \left\langle C_1, ..., C_{|G|} \right\rangle \mid \forall i \in \{1, ..., |G|\}, C_i \in CS \text{ and } |C_i| = g_i \text{ and } \forall j \in \{1, ..., |G|\} : j \neq i, C_i \cap C_j = \emptyset \right\} \tag{8}$$

For example, $\hat{T}_{[1,1,2]}^{\{\{a_1\},\{a_2\},\{a_3,a_4\}\}} = \{\langle \{a_1\}, \{a_2\}, \{a_3, a_4\} \rangle, \langle \{a_2\}, \{a_1\}, \{a_3, a_4\} \rangle\}$. Next, given any coalition structure $CS \in P_G$, we will prove that $|\hat{T}_G^{CS}| \geq 1$. To this end, let $\left\langle C_1, C_2, \ldots, C_{|G|} \right\rangle$ be an ordering on the coalitions that belong to $CS$. Then, from (7) and (8), we can see that:

$$\left\langle C_1, C_2, \ldots, C_{|G|} \right\rangle \in \hat{T}_G^{CS} \quad \text{iff} \quad |C_1| \leq |C_2| \leq \ldots \leq \left| C_{|G|} \right| \tag{9}$$

Now since there is at least one way of ordering the coalitions in $CS$ such that $|C_1| \leq ... \leq \left| C_{|G|} \right|$, then there is at least one ordered set in $\hat{T}_G^{CS}$. In other words, $|\hat{T}_G^{CS}| \geq 1$. This, in turn, implies that every coalition structure in $P_G$ is generated by MCP.

□

## Appendix E. Proof of Theorem 4.

Given an integer partition $G = [g_1, \ldots, g_{|G|}] \in \mathcal{G}$, let $\widetilde{T}_G$ be the set of ordered sets that are generated by FCP. Moreover, given a coalition structure $CS \in P_G$, let $\widetilde{T}_G^{CS}$ be a subset of $\widetilde{T}_G$ containing all the ordered sets that correspond to $CS$. That is:

$$\widetilde{T}_G^{CS} = \left\{ \left\langle C_1, C_2, \ldots, C_{|G|} \right\rangle \in \widetilde{T}_G \mid \forall i \in \{1, 2, \ldots, |G|\}, C_i \in CS \right\}$$

Next, we will prove that $|\widetilde{T}_G^{CS}| = 1$. We define $\hat{T}_G$ and $\hat{T}_G^{CS}$ as in Appendix D. We also assume, without loss of generality, that the order in (7) holds. Note that, if $G(g_i) = 1 \; \forall i \in \{1, \ldots, |G|\}$, then there is no difference between the way FCP works and the way MCP works.[27] On the other hand, if there exists $i \in \{1, ..., |G|\}$ such that $G(g_i) > 1$, then the only difference between FCP and MCP is that FCP avoids some of the coalition structures that are generated by MCP. This implies

26. Recall that $T_G$ is the cartesian product of the lists: $L_s : s \in G$.

27. Recall that $G(g_i)$ is the multiplicity of $g_i$ in $G$.

that:

$$\text{if } G(g_i) = 1\ \forall i \in \{1, \ldots, |G|\} \quad \text{then} \quad \widetilde{T}_G = \hat{T}_G \ \text{ and } \ \forall CS \in P_G, \widetilde{T}_G^{CS} = \hat{T}_G^{CS} \tag{10}$$

$$\text{else} \quad \widetilde{T}_G \subseteq \hat{T}_G \ \text{ and } \ \forall CS \in P_G, \widetilde{T}_G^{CS} \subseteq \hat{T}_G^{CS} \tag{11}$$

Now, given a coalition structure $CS \in P_G$, let $\langle C_1, C_2, \ldots, C_{|G|} \rangle$ be defined as in Appendix D (i.e. it is an ordering on the coalitions that belong to $CS$). Then, from (9), we find that the number of ordered sets in $\hat{T}_G^{CS}$ is equal to the number of possible ways of ordering the coalitions in $CS$ such that: $|C_1| \leq |C_2| \leq ... \leq |C_{|G|}|$. Based on this, as well as (10) and (11), we distinguish between two cases:

- If $G(g_i) = 1\ \forall i \in \{1, ..., |G|\}$, then there would only be one possible way of ordering the coalitions in $CS$ such that $|C_1| \leq ... \leq |C_{|G|}|$ (because every coalition in $CS$ has a unique size). This implies that $|\hat{T}_G^{CS}| = 1$, and from (10), we find that $|\widetilde{T}_G^{CS}| = 1$.
- If $\exists i \in \{1, ..., |G|\} : G(g_i) > 1$, then there would be multiple ways of ordering the coalitions in $CS$ such that $|C_1| \leq ... \leq |C_{|G|}|$, which implies that $|\hat{T}_G^{CS}| > 1$. However, from (11), we know that $\widetilde{T}_G^{CS}$ is a subset of $\hat{T}_G^{CS}$. Then, by proving that $\widetilde{T}_G^{CS}$ contains exactly one of the ordered sets in $\hat{T}_G^{CS}$, we prove that $|\widetilde{T}_G^{CS}| = 1$. To be more precise, in case we have: $|C_x| = |C_{x+1}| = ... = |C_{x+y}|$, then every possible permutation of those coalitions will be generated by MCP, and we need to prove that only one of them will be generated by FCP. Based on this, if we denote by $c_k$ the smallest[28] agent in $C_k$, then it is sufficient to prove that FCP only generates the one permutation that satisfies: $c_x < c_{x+1} < ... < c_{x+y}$. Note that the agents in $A_k$ are ordered such that $A_{k,1} < A_{k,2} < \cdots < A_{k,|A_k|}$. Based on this, if $c_k = A_{k,i}$, then there are $i - 1$ agents in $A_k$ that are smaller than $c_k$, and since $A_{k+1} = A_k \backslash C_k$, then there are $i - 1$ agents in $A_{k+1}$ that are smaller than $c_k$. Therefore, to ensure that $c_k < c_{k+1}$, it is sufficient to generate $C_{k+1}$ such that it does not contain the first (i.e. smallest) $i - 1$ agents of $A_{k+1}$. For example, given $A_k = \langle a_1, a_4, a_5, a_7, a_8, a_9 \rangle$ and $M_k = \{3, 5\}$, we would have $C_k = \{a_5, a_8\}$ and $c_k = A_{k,3}$. This implies that $A_{k+1}$ contains two agents that are smaller than $c_k$ (namely, agents $a_1$ and $a_4$). Therefore, to ensure that $c_k < c_{k+1}$, it is sufficient to generate $C_{k+1}$ such that it does not contain the first (i.e. smallest) two agents in $A_{k+1}$. This can be done by ensuring that $M_{k+1}$ does not contain elements 1 or 2. In other words, it can be done by ensuring that $M_{k+1,1} \geq M_{k,1}$, which is a direct result of the way FCP is modified.

By proving that $|\widetilde{T}_G^{CS}| = 1$ for all $CS \in P_G$, we prove that FCP generates every coalition structure in $P_G$ exactly once.

□

28. Recall that, for any two agents $a_i, a_j \in A$, we say that $a_i$ is smaller than $a_j$ if and only if $i < j$. This comes from the assumed ordering over the set of agents (see Section 2 for more detail).

## Appendix F. Proof of Theorem 5

We will first prove Theorem 5 for the normal distribution case (i.e. the case where $\forall C \subseteq A, v(C) \sim |C| \times N(\mu, \sigma^2)$). Specifically, we will show how the coalition structures that contain fewer coalitions are more likely to be optimal. In order to prove this, we will first prove the following lemma which deals with properties of the normal distribution.

**Lemma 1.** *For any given value $r \in \mathbb{R}$, and for any two random variables $X_a \sim N(\mu, {\sigma_a}^2)$ and $X_b \sim N(\mu, {\sigma_b}^2)$ such that $\sigma_a < \sigma_b$, the following holds:*

$$P(X_a > r) < P(X_b > r) \tag{12}$$

**Proof.** Given $r \in \mathbb{R}$, let $\Phi_{\mu,\sigma_a^2}(r)$ and $\Phi_{\mu,\sigma_b^2}(r)$ be the cumulative distribution functions of $N(\mu, {\sigma_a}^2)$ and $N(\mu, {\sigma_b}^2)$ respectively. That is,

$$\Phi_{\mu,\sigma_a^2}(r) = \frac{1}{2}\left(1 + \text{erf}\left(\frac{r-\mu}{\sigma_a\sqrt{2}}\right)\right)$$

$$\Phi_{\mu,\sigma_b^2}(r) = \frac{1}{2}\left(1 + \text{erf}\left(\frac{r-\mu}{\sigma_b\sqrt{2}}\right)\right)$$

where $\text{erf}(M) = \frac{2}{\sqrt{\pi}}\int_0^M e^{-t^2}$ is the error function. Then, in order to prove that the inequality in (12) holds, it is sufficient to prove that:

$$\Phi_{\mu,\sigma_a^2}(r) > \Phi_{\mu,\sigma_b^2}(r) \tag{13}$$

To this end, given that $\sigma_a < \sigma_b$, the following holds, where $abs(M)$ is the absolute value of $M$:

$$abs\left(\frac{r-\mu}{\sigma_a\sqrt{2}}\right) > abs\left(\frac{r-\mu}{\sigma_b\sqrt{2}}\right)$$

This, in turn, implies that:

$$\text{erf}\left(\frac{r-\mu}{\sigma_a\sqrt{2}}\right) > \text{erf}\left(\frac{r-\mu}{\sigma_b\sqrt{2}}\right)$$

Based on this, as well as the fact that $\text{erf}(M) \geq 0$, we deduce that (13) holds.

□

Based on the above lemma, and given a coalition structure $CS' : |CS'| > 1$, we will prove that there exists another coalition structure $CS'' : |CS''| < |CS'|$ such that:

$$P\left(V(CS'') = V(CS^*)\right) > P\left(V(CS') = V(CS^*)\right)$$

In more detail, let $CS' = \{C_{x_1}, \cdots, C_{x_\alpha}, C_{y_1}, \cdots, C_{y_\beta}\}$ and $CS'' = \{C_x, C_{y_1}, \cdots, C_{y_\beta}\}$ such that $C_x = C_{x_1} \cup \cdots \cup C_{x_\alpha}$. Then, based on the properties of the normal distribution, we have:

$$v(C_x) \sim N\left(|C_x| \times \mu, |C_x|^2 \times \sigma^2\right) \tag{14}$$

and:

$$v(C_{x_1}) + \cdots + v(C_{x_\alpha}) \sim N\Big((|C_{x_1}| + \cdots + |C_{x_\alpha}|) \times \mu, (|C_{x_1}|^2 + \cdots + |C_{x_\alpha}|^2) \times \sigma^2\Big) \quad (15)$$

Now, given a coalition structure $CS$, let $\mu_{CS}$ and $\sigma^2_{CS}$ denote the mean and variance of the distribution of $V(CS)$. Then, based on (14) and (15), we have:

$$\mu_{CS''} = (|C_x| \times \mu) + \sum_{C \in CS'' \setminus \{C_x\}} (|C| \times \mu)$$

$$\sigma^2_{CS''} = (|C_x|^2 \times \sigma^2) + \sum_{C \in CS'' \setminus \{C_x\}} (|C| \times \sigma)^2$$

and we have:

$$\mu_{CS'} = ((|C_{x_1}| + \cdots + |C_{x_\alpha}|) \times \mu) + \sum_{C \in CS' \setminus \{C_{x_1}, \cdots, C_{x_\alpha}\}} (|C| \times \mu)$$

$$\sigma^2_{CS'} = \Big((|C_{x_1}|^2 + \cdots + |C_{x_\alpha}|^2) \times \sigma^2\Big) + \sum_{C \in CS' \setminus \{C_{x_1}, \cdots, C_{x_\alpha}\}} (|C| \times \sigma)^2$$

Since $|C_x| = |C_{x_1}| + \cdots + |C_{x_\alpha}|$, and since $CS'' \setminus \{C_x\} = CS' \setminus \{C_{x_1}, \cdots, C_{x_\alpha}\}$, we can see that the distribution of $V(CS'')$ and $V(CS')$ only differ by the way their variances differ. Note that:

$$|C_x|^2 = (|C_{x_1}| + \cdots + |C_{x_\alpha}|)^2 > |C_{x_1}|^2 + \cdots + |C_{x_\alpha}|^2$$

This implies that $\sigma^2_{CS'} < \sigma^2_{CS''}$. Therefore, based on Lemma 1, we find that for any value $r \in \mathbb{R}$:

$$P(V(CS') > r) < P(V(CS'') > r)$$

In other words, it is more likely for $CS''$ to have a value greater than $r$, which implies that it is more likely for $CS''$ to be the optimal coalition structure.

Having proved Theorem 5 for the normal distribution case, we will now give the intuition behind the proof for the uniform distribution case (i.e. the case where $\forall C \subseteq A, v(C) \sim |C| \times U(a,b)$). Specifically, assuming that $CS'$ and $CS''$ are defined as above, we would have $v(C_x) \sim |C_x| \times U(a,b)$ and, for any coalition $C \in \{C_{x_1}, \cdots, C_{x_\alpha}\}$, we would have $v(C) \sim |C| \times U(a,b)$. Then, it is easy to verify that $P(v(C_x) \leq r)$ is less than $P(v(C_{x_1}) + \cdots + v(C_{x_\alpha}) \leq r)$ for high values of $r$. The intuition behind this difference in probabilities is that *the sum of* Uniformly distributed variables (called a Uniform Sum distribution) results in a distribution giving lower probability to low and high values, and higher probability to middle ranged values. Instead, for a uniformly distributed variable, all values are equally probable. Therefore, given a Uniform Sum distribution and a Uniform distribution with the same minimum and maximum values, the Uniform distribution will give a higher probability to higher values. Hence, the above proof holds for the Uniform distribution as well.

□

## Appendix G. Proof of Theorem 6.

Given the following:

$$\forall C \subseteq A,\ v(C) \sim N(|C|, |C|) \tag{16}$$

we need to prove that the value of every coalition structure is independently drawn from the same normal distribution. Specifically, we will prove that the following holds:

$$\forall CS \in \mathcal{P},\ V(CS) \sim N(|A|, |A|) \tag{17}$$

From the properties of the normal distribution, we know that, for any two independent random variables, $x$ and $y$ such that $x \sim N(\mu_x, {\sigma_x}^2)$ and $y \sim N(\mu_y, {\sigma_y}^2)$, we have:

$$(x + y) \sim N(\mu_x + \mu_y, {\sigma_x}^2 + {\sigma_y}^2) \tag{18}$$

Then, based on (16) and (18), any two coalition values, $v(C_1)$ and $v(C_2)$, satisfy the following (since they are independent random variables):

$$(v(C_1) + v(C_2)) \sim N(|C_1| + |C_2|, |C_1| + |C_2|)$$

This implies that the following is true:

$$\forall CS \in \mathcal{P},\ \left(\sum_{C \in CS} v(C)\right) \sim N\left(\sum_{C \in CS} |C|, \sum_{C \in CS} |C|\right) \tag{19}$$

Finally, note that we assume the following:

$$\forall CS \in \mathcal{P},\ V(CS) = \sum_{C \in CS} v(C) \tag{20}$$

$$\forall CS \in \mathcal{P},\ \forall C, C' \in CS,\ C \cap C' = \emptyset \tag{21}$$

Then, from (19), (20), and (21), we find that:

$$\forall CS \in \mathcal{P},\ V(CS) \sim N(|\cup_{C \in CS}|, |\cup_{C \in CS}|)$$

which implies that (17) holds since $\cup_{C \in CS} = A$.

□